\documentclass[12pt]{article}

\pdfoutput=1
\usepackage[dvipsnames]{xcolor}
\usepackage{amsfonts}
\usepackage{mathtools}
\usepackage{amsmath}
\usepackage{latexsym}
\usepackage{graphicx}
\usepackage{color}
\usepackage{amssymb}
\usepackage{braket}
\usepackage[many]{tcolorbox}
\numberwithin{equation}{section}
\usepackage{tikz}
\usetikzlibrary{intersections, calc}
\usetikzlibrary {arrows.meta,bending,positioning}
\usepackage[hang,small,bf]{caption}
\usepackage[subrefformat=parens]{subcaption}
\allowdisplaybreaks
\usepackage{afterpage}
\usepackage{slashed}
\everymath{\displaystyle}
\usepackage{url}
\usepackage{cite}
\usepackage{bm}
\usepackage{cancel}
\usepackage{titling}
\usepackage{lmodern}
\usepackage{hyperref}

\hypersetup{colorlinks,linkcolor=BrickRed,citecolor=BrickRed,urlcolor=BrickRed}

\usepackage{comment}

\def\d{\mathsf{d}}
\def\i{\text{i}}
\renewcommand{\thefootnote}{\fnsymbol{footnote}}

\def\CL{\mathcal{L}}

\newcommand{\Sw}{\widetilde{S}}

\pretitle{%
  \begin{flushright}
    \small OU-HET 1270
  \end{flushright}
  \begin{center}
  \LARGE
}
\posttitle{\end{center}}
\preauthor{\begin{center} \large}
\postauthor{\end{center}}
\predate{\begin{center} \small}
\postdate{\end{center}}

\title{\vspace{1cm} Boundary Scattering and Non-invertible Symmetries \\ in $1+1$ Dimensions \vspace{2cm}}

\author{Soichiro Shimamori and Satoshi Yamaguchi \\ \vspace{5mm}
{\small\it Department of Physics, Osaka University, Machikaneyama-Cho 1-1, Toyonaka 560-0043, Japan}}

\date{April 2025}

\begin{document}
\maketitle
\thispagestyle{empty}

\begin{abstract}
Recent studies by Copetti, C\'ordova and Komatsu have revealed that when non-invertible symmetries are spontaneously broken, the conventional crossing relation of the S-matrix is modified by the effects of the corresponding topological quantum field theory (TQFT). In this paper, we extend these considerations to 
$(1+1)$-dimensional quantum field theories (QFTs) with boundaries. In the presence of a boundary, one can define not only the bulk S-matrix but also the boundary S-matrix, which is subject to a consistency condition known as the boundary crossing relation. We show that when the boundary is weakly-symmetric under the non-invertible symmetry, the conventional boundary crossing relation also receives a modification due to the TQFT effects. As a concrete example of the boundary scattering, we analyze kink scattering in the gapped theory obtained from the 
$\Phi_{(1,3)}$-deformation of a minimal model. We explicitly construct the boundary S-matrix that satisfies the Ward-Takahashi identities associated with non-invertible symmetries.
\end{abstract}

\renewcommand{\thefootnote}{\arabic{footnote}}
\setcounter{footnote}{0}

\newpage
\pagenumbering{arabic}
\setcounter{page}{1}
\tableofcontents
\section{Introduction}
Scattering amplitudes in $1+1$ dimensions have long and rich histories, serving as key ingredients in our understanding of non-perturbative dynamics of quantum field theories (QFTs). In particular, when a QFT possesses an integrability, i.e., an infinite number of integrals of motion, the S-matrix factorizes into the product of purely elastic two-body scattering matrices. This factorization property immediately leads to the Yang-Baxter equation. By combining this with the unitarity condition and crossing relation, we are able to non-perturbatively determine scattering amplitudes, which provides non-trivial predictions on e.g., the mass spectrum of the gapped theories \cite{Zamolodchikov:1977nu, Shankar:1977cm, Zamolodchikov:1978xm,  Parke:1980ki, Zamolodchikov:1989hfa}.

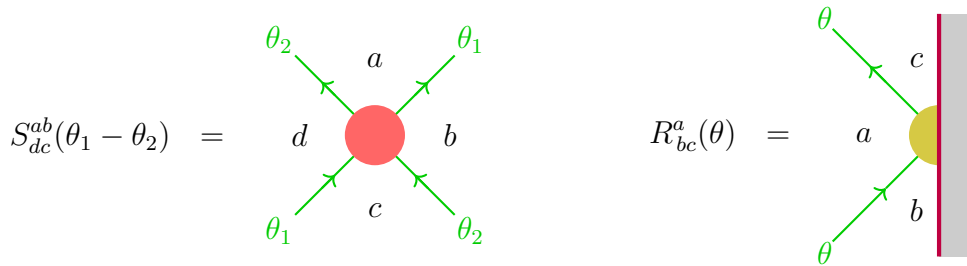
\begin{figure}
    \centering
        \begin{tikzpicture}
        \begin{scope}
                   \node[right] at (-1,0) {$S^{ab}_{dc}(\theta_{1}-\theta_{2})\ \ =$}; 
        \foreach \angle in {225, 315} {
             \draw[thick, black!20!green] (4,0) -- ++(\angle:1);
         }
         \draw[thick, black!20!green, ->] ($(4,0) + (225:1.5)$) -- ++(45:0.75);
         \draw[thick, black!20!green, ->] ($(4,0) + (315:1.5)$) -- ++(135:0.75);
        \foreach \angle in {45, 135} {
             \draw[thick, black!20!green, ->] (4,0) -- ++(\angle:1);
             \draw[thick, black!20!green] ($(4,0) + (\angle:0.75)$) -- ++(\angle:0.75);
         }
        \node [black!20!green] at ($(4,0) + (225:1.8)$) {$\theta_{1}$};
        \node [black!20!green] at ($(4,0) + (315:1.8)$) {$\theta_{2}$};
        \node [black!20!green] at ($(4,0) + (45:1.8)$) {$\theta_{1}$};
        \node [black!20!green] at ($(4,0) + (135:1.8)$) {$\theta_{2}$};
        \fill [red!60] (4,0) circle (0.4);
        \node at ($(4,0) + (0:1)$) {$b$}; 
        \node at ($(4,0) + (90:1)$) {$a$}; 
        \node at ($(4,0) + (180:1)$) {$d$}; 
        \node at ($(4,0) + (270:1)$) {$c$};
        \end{scope}
        \begin{scope}[shift={(7.5cm, 0)}]
                   \node[right] at (0,0) {$R^{a}_{\, bc}(\theta)\ \ =$}; 
        \foreach \angle in {225} {
             \draw[thick, black!20!green] (4,0) -- ++(\angle:1);
         }
         \draw[thick, black!20!green, ->] ($(4,0) + (225:2)$) -- ++(45:1);
        \foreach \angle in {135} {
             \draw[thick, black!20!green, ->] (4,0) -- ++(\angle:1.3);
             \draw[thick, black!20!green] ($(4,0) + (\angle:1)$) -- ++(\angle:1);
         }
        \node [black!20!green, below left] at ($(4,0) + (225:1.8)$) {$\theta$};
        \node [black!20!green, above left] at ($(4,0) + (135:1.8)$) {$\theta$};
        \fill [black!20!yellow] (4,0) circle (0.4);
        \node at ($(3.7,1)$) {$c$}; 
        \node at ($(4,0) + (180:1)$) {$a$}; 
        \node at ($(3.7,-1)$) {$b$};
        \fill[black!20!white] (4, 1.614)--(4.5, 1.614)--(4.5, -1.614)--(4, -1.614);
        \draw[purple, ultra thick] (4, 1.614)--(4, -1.614);
        \end{scope}
\end{tikzpicture}
    \caption{Scattering data in quantum field theories with boundaries. The left panel represents the bulk S-matrix, while the right panel does the boundary S-matrix. The green line means the worldline of a kink, and the red line represents the boundary. The time direction flows from bottom to top.}
    \label{fig:boundary_scattering}
\end{figure}

The crossing relation is a fundamental property of the S-matrix, reflecting the equivalence between different scattering processes related by analytic continuation. It allows one to relate the scattering amplitude in the $s$-channel to the one in the $t$-channel. In the case of the kink scattering in $1+1$ dimensions, the following crossing relation has been historically adopted:
\begin{align}\label{eq:crossing_relation_conv_intro}
            S^{ab}_{dc}(\theta)\overset{?}{=}S^{bc}_{ad}(\i\pi-\theta)\ ,
\end{align}
where $S^{ab}_{dc}(\theta)$ is the S-matrix corresponding to the two-body scattering. The Latin indices correspond to the vacuum indices separated by the worldlines of kinks, and $\theta$ is the relative rapidity between the initial two kinks. (See the left panel in Fig.\ref{fig:boundary_scattering}. For further details, see the main text.)

Recently, Copetti, C\'ordova and Komatsu \cite{Copetti:2024rqj, Copetti:2024dcz} have revisited the scattering amplitudes of the gapped theories arising from relevant deformations that preserve certain non-invertible symmetries of the UV theories. Unlike conventional group-like symmetries, non-invertible symmetries are subject to the categorical fusion rules. See, for instance, \cite{Verlinde:1988sn, Frohlich:2004ef, Frohlich:2006ch, Frohlich:2009gb, Bhardwaj:2017xup, Chang:2018iay,  Komargodski:2020mxz, Thorngren:2021yso, Huang:2021zvu} for early works on non-invertible symmetries in two-dimensional field theories\footnote{If the readers are interested in the non-invertible symmetries in higher dimensions, we recommend e.g., \cite{Shao:2023gho, Schafer-Nameki:2023jdn} and references therein.}. In gapped theories where non-invertible symmetries are spontaneously broken, kinks are generally associated with the twisted sectors of non-invertible symmetry defects\footnote{While the appearance of breathers is certainly possible, we do not consider this possibility in the present work. See \cite{Cordova:2024vsq, Cordova:2024iti, Copetti:2024dcz} for the discussions on the non-invertible symmetries and the breather.}. The dynamics of the corresponding topological quantum field theories (TQFTs) modify the normalization of the S-matrix, which had been widely believed to be correct. As a result, it turns out that the conventional crossing relation \eqref{eq:crossing_relation_conv_intro} needs to be modified by a TQFT factor as follows \cite{Copetti:2024rqj, Copetti:2024dcz}:
\begin{align}\label{eq:crossing_relation_new_intro}
    {S}^{ab}_{dc}(\theta)=\sqrt{\frac{\d_{a} \d_{c}}{\d_{b} \d_{d}}}\, {S}^{bc}_{ad}(\i\pi-\theta)\ ,
\end{align}
where $\d_{a}$ is the quantum dimension of the topological line defect $\mathcal{L}_{a}$ associated to the vacuum index $a$\footnote{Throughout this paper, we assume that the vacuum state $\ket{a}$ is an object of the regular module category with respect to the non-invertible symmetry. For the modified crossing relation in the case of the non-regular module category, see \cite{Copetti:2024dcz}.}.

In this work, we extend these discussions to $(1+1)$-dimensional QFTs with boundaries. In the presence of a boundary, one can introduce not only the bulk S-matrix but also the {\it boundary S-matrix} which describes the reflection of solitons and particles from the boundary~\cite{Ghoshal:1993tm}. As will be reviewed later, the boundary S-matrix must satisfy several consistency conditions, just as the bulk S-matrix does \cite{Ghoshal:1993tm}. In particular, as a boundary analog of the crossing relation \eqref{eq:crossing_relation_conv_intro}, Ghoshal and Zamolodchikov proposed the following boundary crossing relation \cite{Ghoshal:1993tm}:
\begin{align}\label{eq:bdy_crossing_relation_intro}
    {R}^{b}_{\, ca}\left(\frac{\i\pi}{2}-\theta\right)&\overset{?}{=}\sum_{d}S^{ba}_{cd}(2\theta)\, {R}^{d}_{\, ca}\left(\frac{\i\pi}{2}+\theta\right)\ ,
\end{align}
where ${R}^{a}_{\, bc}(\theta)$ is the boundary S-matrix as described in the right panel in Fig.\ref{fig:boundary_scattering}. We should remark that the bulk integrability is implicitly assumed in the boundary crossing relation.  Throughout this paper, we assume that the boundary $\mathcal{B}$ is weakly-symmetric under the bulk non-invertible symmetries, i.e., the topological junction space $\text{Hom}(\mathcal{L}_{a}\otimes\mathcal{B}, \mathcal{B})$ is non-trivial for all $a$ \cite{Choi:2023xjw}\footnote{For recent developments on non-invertible symmetries in boundary QFTs, see also \cite{Aasen:2020jwb, Koide:2023rqd, Putrov:2024uor, Copetti:2024onh, Choi:2024tri, Bhardwaj:2024igy, Choi:2024wfm, Antinucci:2024izg}.}. Then, we find that the above conventional boundary crossing relation \eqref{eq:bdy_crossing_relation_intro} should be modified as 
\begin{center}
\begin{tcolorbox}[colframe=cyan!30!white, colback=cyan!5!white, sharp corners=south, boxrule=0.8mm, width=0.9\textwidth, coltitle=black, breakable]
\begin{equation}\label{eq_modified_boundary_crossing_intro}
    {R}^{b}_{\, ca}\left(\frac{\i\pi}{2}-\theta\right)=\sum_{d}\sqrt{\frac{\d_d}{\d_b}}\, {S}^{ba}_{cd}(2\theta)\, {R}^{d}_{\, ca}\left(\frac{\i\pi}{2}+\theta\right)\ .
\end{equation}
\end{tcolorbox}
\end{center}
This is the main result of this paper, and its derivation will be discussed in detail in the main text.

As a concrete example of the boundary scattering, we investigate the gapped theory where the UV boundary CFT is given by the minimal model $M_p$ labeled by an integer $p\geq 4$, and deformed by the bulk primary operator $\Phi_{(1,3)}$. This deformation does not spoil the integrability in UV boundary CFT, hence the bulk and boundary S-matrices must be subject to the Yang-Baxter equations \cite{Ghoshal:1993tm} in addition to unitarity and (modified) crossing relations. Also, the fusion categorical symmetry $\mathcal{A}_p$ in the UV BCFT is preserved along with this bulk renormalization group (RG) flow. The topological nature of symmetry defects immediately leads to the $(p-1)$-degenerate vacua \cite{Chang:2018iay}.
As mentioned above, we assume that the boundary is weakly-symmetric under these Verlinde lines. This implies that the boundary S-matrices must enjoy the Ward-Takahashi identities associated with these topological lines. As a consequence of non-invertible symmetries, we show that the non-diagonal scatterings are prohibited (i.e., $R^{a}_{\, bc}=0$ for $b\not= c$) whereas the diagonal scatterings are allowed for $p\geq5$. We also explicitly derive the boundary S-matrix by particularly using the modified boundary crossing relation~\eqref{eq_modified_boundary_crossing_intro}. 

This paper is structured as follows. In Sec.~\ref{sec:bulk_scattering}, we summarize fundamental aspects of bulk scattering, serving also as setting the notation. By carefully examining the normalization of the S-matrix, we rederive the modified crossing relation~\eqref{eq:crossing_relation_new_intro} for the bulk S-matrix. In Sec.~\ref{sec:boundary_scattering}, we begin with a review of boundary scattering following Ghoshal and Zamolodchikov. We then proceed to derive the modified boundary crossing relation~\eqref{eq_modified_boundary_crossing_intro}. In Sec.~\ref{sec:boundary_scattering_example}, we present explicit examples of boundary scattering as described above. In particular, we demonstrate how the Ward-Takahashi identities associated to non-invertible symmetries impose strong constraints on boundary scattering. In Sec.~\ref{sec:conclusion},  we conclude with a summary of this work and discuss potential future directions.

\section{Bulk Scattering Amplitudes in $1+1$ dimensions}\label{sec:bulk_scattering}
In this section, we review some fundamental aspects of the scattering amplitudes in 1+1 dimensions without boundaries. In Sec.~\ref{subsec:bulk_scat1}, we begin by introducing several physical quantities that characterize the bulk scattering among kinks, followed by a discussion on the unitarity condition, Yang-Baxter equation and historically accepted crossing relation. In Sec.~\ref{subsec:modified_crossing_bulk}, we elaborate on how the crossing relation should be modified by the TQFT dynamics when the non-invertible symmetries are spontaneously broken in the gapped theories. Finally, as an example, we consider the scattering amplitude of the gapped theory obtained by the RG flow from the UV minimal model $M_{p}$ $(p\geq 4)$ via the $\Phi_{(1,3)}$-deformation, which satisfies the modified crossing relation.
\subsection{Bulk S-matrix}\label{subsec:bulk_scat1}  
We first describe some key concepts related to the scattering amplitudes in the gapped theories in the absence of boundaries. Suppose that a UV completion of the gapped theory possesses the categorical symmetry\footnote{In this paper, we use the terms categorical symmetry and non-invertible symmetry interchangeably.} $\mathcal{A}_p$ where the simple topological lines $\mathcal{L}_a\, ({a=1, 2, \cdots, p-1})$ are subject to the multiplicity-free fusion rule:
\begin{align}\label{fusion_rule}
    \mathcal{L}_a \otimes \mathcal{L}_b = \sum_{c}N_{\, bc}^{a} \, \mathcal{L}_c\ ,  
\end{align}
where $N_{\, bc}^{a}$ is given by 
\begin{align}
    \begin{aligned}
        N_{\, bc}^{a}=
        \begin{cases}
            1\ , \quad \left(1+|a-b| \leq c\leq \min(a+b-1, 2p-1-a-b) \ \text{and}\ c+a+b=1\mod 2\right)\ , \\
            0\ , \quad (\text{otherwise})\ . 
        \end{cases}
    \end{aligned}
\end{align}
This fusion rule defines the trivalent junction among three topological defects $\CL_a$, $\CL_b$ and $\CL_c$ when $N^{a}_{\, bc}=1$:\footnote{In the case of the fusion category $\mathcal{A}_{p}$, we do not need to care about the orientation of the topological line. This is because the topological line $\mathcal{L}_{a}$ serves as its own dual, as indicated by $\text{dimHom}(\mathcal{L}_{a}\otimes\mathcal{L}_{a}, \mathcal{L}_{1})=1$.}
\begin{align}
\begin{tikzpicture}[scale=0.7, baseline=0cm]
    \filldraw [blue!80] (0,0) circle (2pt);
    \draw[thick, blue!80] (0,0)--(0, 0.75);
    \draw[thick, blue!80] (0, 0.75)-- (0, 1.5) node[above=0.05cm] {$\CL_c$};
    \draw[thick, blue] ({1.5*cos(210)}, {1.5*sin(210)}) node[below left] {$\CL_a$} -- ({0.75*cos(210)}, {0.75*sin(210)});
    \draw[thick, blue] ({0.75*cos(210)}, {0.75*sin(210)}) -- (0,0);
    \draw[thick, blue] ({1.5*cos(-30)}, {1.5*sin(-30)}) node[below right] {$\CL_b$} -- ({0.75*cos(-30)}, {0.75*sin(-30)});
    \draw[thick, blue] ({0.75*cos(-30)}, {0.75*sin(-30)}) -- (0,0);
\end{tikzpicture}
\end{align}
The following fusion property becomes particularly important for our purpose:
\begin{align}\label{eq:fusionproperty}
    \begin{tikzpicture}[scale=0.8, baseline=0cm]
        \begin{scope}[shift={(-0.5cm, 0)}]
        \draw[thick, blue!80] (10,-1)node[below] {$\mathcal{L}_{a}$} -- (10, 0.5);
        \draw[thick, blue!80] (10, 0.5) -- (10, 2);
        \draw[thick, blue!80] (12,-1) node[below] {$\mathcal{L}_{b}$} -- (12, 0.5);
        \draw[thick, blue!80] (12, 0.5) -- (12, 2);
        \end{scope}
        \node at (14, 0) {{$=\ \ \sum\limits_{c}\ \ \sqrt{\frac{\d_{c}^2 }{\d_a \d_b}}$}};
        \begin{scope}[shift={(18cm, -0.5cm)}]
                \filldraw [blue!80] (0,0) circle (2pt);
    \draw[thick, blue!80] (0,0)--(0, 0.75);
    \draw[thick, blue!80] (0, 0.75)-- (0, 1.5);
    \draw[thick, blue] ({1.5*cos(210)}, {1.5*sin(210)}) node[below left] {$\CL_a$} -- ({0.75*cos(210)}, {0.75*sin(210)});
    \draw[thick, blue] ({0.75*cos(210)}, {0.75*sin(210)}) -- (0,0);
    \draw[thick, blue] ({1.5*cos(-30)}, {1.5*sin(-30)}) node[below right] {$\CL_b$} -- ({0.75*cos(-30)}, {0.75*sin(-30)});
    \draw[thick, blue] ({0.75*cos(-30)}, {0.75*sin(-30)}) -- (0,0);
    \node[blue!80] at (0.5, 0.75) {$\CL_c$};
    \draw[thick, blue] (0,1.5)-- ++ ({0.8*cos(150)}, {0.8*sin(150)});
    \draw[thick, blue] ({0.8*cos(150)}, {1.5+0.8*sin(150)})-- ++ ({0.75*cos(150)}, {0.75*sin(150)}) node[above left] {$\CL_a$};
    \draw[thick, blue] (0,1.5)-- ++ ({0.8*cos(30)}, {0.8*sin(30)});
    \draw[thick, blue] ({0.8*cos(30)}, {1.5+0.8*sin(30)})-- ++ ({0.75*cos(30)}, {0.75*sin(30)}) node[above right] {$\CL_b$};
    \filldraw [blue!80] (0,1.5) circle (2pt);
        \end{scope}
    \end{tikzpicture}
\end{align}
We also assume that this categorical symmetry $\mathcal{A}_p$ is preserved along with the RG flow, and spontaneously broken in the IR regime. Then, the vacuum states can be characterized by the module category of $\mathcal{A}_p$ in general \cite{Choi:2023xjw}. For simplicity, we only consider the case where these vacua are objects of the {\it regular module category}; hence, we can express a vacuum state $\ket{a}$ as
\begin{align}\label{eq:vacuum_states}
    \ket{a}=\mathcal{L}_{a}\ket{\tilde{0}}\ , 
\end{align}
where $\ket{\tilde{0}}$ is the reference vacuum that corresponds to the identity line $\mathcal{L}_{1}=\bm{1}$. Since these vacua are degenerate thanks to the categorical symmetry $\mathcal{A}_p$ \cite{Chang:2018iay}, we can introduce kink solitons with identical mass that interpolate between adjacent vacua in the gapped theory. We denote the kink state by $\ket{K_{ab}(\theta)}$, which is characterized by the rapidity $\theta$ and the vacua $\ket{a}$ and $\ket{b}$ with $|a-b|=1$. Pictorially, we express the kink state $\ket{K_{ab}(\theta)}$ as follows:
\begin{align}
    \begin{tikzpicture}[baseline=0cm]
         \node[right] at (0,0) {$\ket{K_{ab}(\theta)}\ \ =  $};
        \draw[->, thick, black!20!green] (4,-1) node[below left=0.05cm, black] {} --(4, 0) node[below=1cm] {$\theta$};
        \draw[thick, black!20!green] (4, 0)-- (4, 1) node[midway, right] {};
        \draw node at (3.25, 0) {$a$};
        \draw node at (4.75, 0) {$b$};
    \end{tikzpicture}
\end{align}
where the arrow means the flow of the momentum. In general, we can define the asymptotic in and out states composed of $N$-kinks:
\begin{align}
    \ket{K_{a_{0} a_{1}}(\theta_{1})K_{a_{1} a_{2}}(\theta_{2})\cdots K_{a_{N-1} a_{N}}(\theta_{N})}_{\text{in}/\text{out}} \ . 
\end{align}
Our particular interest here is the $2\to 2$ scattering:
\begin{align}
    K_{ac}(\theta_{1})+K_{c b}(\theta_{2}) \longrightarrow K_{a d}(\theta_{2})+K_{d b}(\theta_{1})\ , 
\end{align}
and the corresponding S-matrix is defined by the unitary transformation from the in to out states as   
\begin{align}\label{eq:2to2}
    \ket{K_{a c}(\theta_{1})K_{c b}(\theta_{2})}_{\text{in}}=S^{db}_{ ac}(\theta_1 , \theta_2 ) \ket{K_{a d}(\theta_{2})K_{d b}(\theta_{1})}_{\text{out}}+\cdots\ .
\end{align}
The Lorentz invariance implies that the S-matrix can only depend on the difference of rapidities $\theta:=\theta_1 -\theta_2$:
\begin{align}
    S^{ab}_{dc}(\theta_1 , \theta_2)=S^{ab}_{dc}(\theta)\ . 
\end{align}
Throughout this paper, we often express this S-matrix as
\begin{align}\label{eq:graphical_S_matrix}
    \begin{tikzpicture}[baseline=0cm]
       \node[right] at (-1,0) {$S^{ab}_{dc}(\theta_1 , \theta_2 )\ \ =$}; 
        \foreach \angle in {225, 315} {
             \draw[thick, black!20!green] (4,0) -- ++(\angle:1);
         }
         \draw[thick, black!20!green, ->] ($(4,0) + (225:1.5)$) -- ++(45:0.75);
         \draw[thick, black!20!green, ->] ($(4,0) + (315:1.5)$) -- ++(135:0.75);
        \foreach \angle in {45, 135} {
             \draw[thick, black!20!green, ->] (4,0) -- ++(\angle:1);
             \draw[thick, black!20!green] ($(4,0) + (\angle:0.75)$) -- ++(\angle:0.75);
         }
        \node [black!20!green] at ($(4,0) + (225:1.8)$) {$\theta_{1}$};
        \node [black!20!green] at ($(4,0) + (315:1.8)$) {$\theta_{2}$};
        \node [black!20!green] at ($(4,0) + (45:1.8)$) {$\theta_{1}$};
        \node [black!20!green] at ($(4,0) + (135:1.8)$) {$\theta_{2}$};
        \fill [red!60] (4,0) circle (0.4);
        \node at ($(4,0) + (0:1)$) {$b$}; 
        \node at ($(4,0) + (90:1)$) {$a$}; 
        \node at ($(4,0) + (180:1)$) {$d$}; 
        \node at ($(4,0) + (270:1)$) {$c$};
\end{tikzpicture}
\end{align}
where we assume that the time direction flows from the bottom to up. 

Determining the S-matrix in an exact manner is generally a challenging problem, however this becomes significantly tractable when the theory possesses the {\it integrability}. In integrable QFTs, there are an infinite number of integrals of motion, which give dramatic constraints on scattering amplitudes. Firstly, scatterings are restricted to purely elastic ones, meaning that no kink or particle production occurs and a set of kink momenta is preserved before and after the scattering. Since there is no particle or kink production, the unitarity condition can be simply put as follows: 
        \begin{align}\label{eq:uni}
            \sum_{b}S^{bc}_{ad}(\theta)\, S^{ec}_{ab}(-\theta)=\delta_{d e}\ ,
        \end{align}
        \begin{align}\label{eq:graphical_unitarity}
    \begin{tikzpicture}[scale=0.8, baseline=0.5cm]
        \node[right] at (0.5,0.8) {{\Large $\sum\limits_{b}$}};
        \foreach \angle in {225, 315} {
             \draw[thick, black!20!green] (4,0) -- ++(\angle:1);
         };
         \draw[thick, black!20!green, ->] ($(4,0) + (225:1.5)$) -- ++(45:0.75);
         \draw[thick, black!20!green, ->] ($(4,0) + (315:1.5)$) -- ++(135:0.75);
        \foreach \angle in {45, 135} {
             \draw[thick, black!20!green, ->] (4,2) -- ++(\angle:1);
             \draw[thick, black!20!green] ($(4,2) + (\angle:0.75)$) -- ++(\angle:0.75);
         };
        \draw[thick, black!20!green][<->] ($(4,1) + (0:0.8)$) arc (0:-180: 0.8cm and 1cm);
        \draw[thick, black!20!green][] ($(4,1) + (0:0.8)$) arc (0:180: 0.8cm and 1cm);
        \node [black!20!green] at ($(4,0) + (225:1.8)$) {$\theta_{1}$};
        \node [black!20!green] at ($(4,0) + (315:1.8)$) {$\theta_{2}$};
        \fill [red!60] (4,0) circle (0.4);
        \fill [red!60] (4,2) circle (0.4);
        \node at ($(4,1) + (0:1.5)$) {$c$}; 
        \node at ($(4,0) + (90:1)$) {$b$}; 
        \node at ($(4,1) + (180:1.5)$) {$a$}; 
        \node at ($(4,0) + (270:1)$) {$d$};
        \node at ($(4,2) + (90:1)$) {$e$};
        \node at (7,1) {{\Large =}};
        \draw[thick, black!20!green, ->] (10,-1.5) -- (10, 0.8);
        \draw[thick, black!20!green] (10, 0.8) -- (10, 3);
        \draw[thick, black!20!green, ->] (12,-1.5) -- (12, 0.8);
        \draw[thick, black!20!green] (12, 0.8) -- (12, 3);
         \draw[thick, black!20!green, ->] ($(4,0) + (315:1.5)$) -- ++(135:0.75);
         \node at ($(11.5,0.5) + (0:1.5)$) {$c$}; 
        \node at ($(11,-0.5) + (90:1)$) {$d$}; 
        \node at ($(10.5,0.5) + (180:1.5)$) {$a$}; 
        \node at (8,1) {{\large $\delta_{de}$}};
\end{tikzpicture}
\end{align}
The second remarkable feature is the so-called {\it factorizability}, meaning that all S-matrices factorize into products of the two-body scattering amplitudes. This factorizable property implies that the S-matrix must be subject to the Yang-Baxter equation      
    \begin{align}\label{eq:YBeq}
        \sum_{g}S^{gd}_{fe}(\theta_1 -\theta_2)\,S^{bc}_{gd}(\theta_1 -\theta_3)\,S^{ab}_{fg}(\theta_2 -\theta_3)=\sum_{g}S^{gc}_{ed}(\theta_2 -\theta_3)\,S^{ag}_{fe}(\theta_1 -\theta_3)\,S^{bc}_{ag}(\theta_1 -\theta_2)\ .
        \end{align}
        \begin{align}
        \begin{tikzpicture}[scale=0.7, baseline=1cm]
        \node[right] at (-8,1.3) {{\Large $\sum\limits_{g}$}};
        \draw[thick, black!20!green] (-6, -1) -- (-1, 4);
        \draw[thick, black!20!green] (-6, 4) -- (-1, -1);
        \draw[thick, black!20!green] (-4.75, -1) -- (-4.75, 4);
        \fill [red!60] (-3.5,1.5) circle (0.4);
        \fill [red!60] (-4.75,0.25) circle (0.4);
        \fill [red!60] (-4.75,2.75) circle (0.4);
        \node at (-5.5, 1.5) {$f$}; 
        \node at (-4.25, 1.5) {$g$};
        \node at (-3.5, 0) {$d$};
        \node at (-5.1, -0.5) {$e$};
        \node at (-2.25, 1.5) {$c$};
        \node at (-3.5, 3) {$b$};
        \node at (-5.1, 3.5) {$a$};
        \node[right] at (0,1.5) {{\Large $=$}};
        \node[right] at (1,1.3) {{\Large $\sum\limits_{g}$}};
        \draw[thick, black!20!green] (8, -1) -- (3, 4);
        \draw[thick, black!20!green] (8, 4) -- (3, -1);
        \draw[thick, black!20!green] (6.75, -1) -- (6.75, 4);
        \fill [red!60] (5.5,1.5) circle (0.4);
        \fill [red!60] (6.75,0.25) circle (0.4);
        \fill [red!60] (6.75,2.75) circle (0.4);
        \node at (7.5, 1.5) {$c$}; 
        \node at (6.25, 1.5) {$g$};
        \node at (5.5, 0) {$e$};
        \node at (7.1, -0.5) {$d$};
        \node at (4.25, 1.5) {$f$};
        \node at (5.5, 3) {$a$};
        \node at (7.1, 3.5) {$b$};
        \end{tikzpicture}
    \end{align}
By combining these two conditions \eqref{eq:uni} and \eqref{eq:YBeq} with the (modified) crossing relation, which will be discussed in the next section, one can drastically constrain the form of the S-matrix in integrable QFTs.
\subsection{Modified Crossing Relation}\label{subsec:modified_crossing_bulk}
In the previous section, we have seen that the S-matrix in an integrable QFT must be subject to unitarity condition and Yang-Baxter equation. In addition to these constraints, it had been historically believed that the S-matrix should satisfy the so-called crossing relation
\begin{align}\label{eq:crossing_relation_conventional}
            S^{ab}_{dc}(\theta)\overset{?}{=}S^{bc}_{ad}(\i\pi-\theta)\ .
\end{align}
However, as pointed out in the papers \cite{Copetti:2024rqj, Copetti:2024dcz}, this crossing relation must be modified when the non-invertible symmetry is spontaneously broken. In this section, we rederive how the above relation \eqref{eq:crossing_relation_conventional} should be corrected due to the TQFT dynamics\footnote{Remark that the discussion of the modified crossing relation applies to QFTs which do not possess the integrability.}.  

To discuss the crossing relation, it is useful to consider the LSZ reduction formula, which relates correlation functions to S-matrix elements.
In the conventional LSZ framework, particles are described by genuine local operators.
However, kink solitons cannot be captured by such operators. This is because for any local operator $\phi(x)$, the matrix element $\braket{K_{ab}(\theta_{2})|\phi(x)|\tilde{0}}$ vanishes, since a genuine local operator cannot change the boundary conditions at spatial infinity.
We can instead utilize non-genuine local operators to describe the kink solitons.
We consider a non-genuine operator $\phi_{ab}(x)$ accompanied by topological line defects $\mathcal{L}_a$ and $\mathcal{L}_b$, depicted as 
\begin{align}
   \phi_{ab}(x)=\ \begin{tikzpicture}[baseline=0cm]
        \draw[thick, blue!80] (2.5, 0) -- (5.5, 0);
        \draw node[below] at (3.25, 0) {$a$};
        \draw node[below] at (4.75, 0) {$b$};
        \fill[black] (4, 0) circle (2pt);
    \end{tikzpicture}
\end{align}
and we assume that $\braket{K_{ab}(\theta_{2})|\phi_{ab}(x)|\tilde{0}}\ne 0$.
We now consider a four-point correlation function of these operators in the reference vacuum $\ket{\tilde{0}}$: 
\begin{align}
    \braket{\phi_{dc}(x_1)\phi_{cb}(x_2)\phi_{ab}(x_3)\phi_{da}(x_4)}=\ 
    \begin{tikzpicture}[baseline=0cm]
       \begin{scope}[scale=0.8]
        \draw [blue!80, thick] (4,0) circle (1.5);
        \node at ($(5.2,0)$) {$b$}; 
        \node at ($(4,1.2)$) {$a$}; 
        \node at ($(4,-1.2)$) {$c$}; 
        \node at ($(2.8,0)$) {$d$}; 
        \fill[black] ($(5.414, 1.414)-(45:0.5)$) circle (2pt);
        \fill[black] ($(5.414, -1.414)-(315:0.5)$) circle (2pt);
        \fill[black] ($(2.586, 1.414)-(135:0.5)$) circle (2pt);
        \fill[black] ($(2.586, -1.414)-(225:0.5)$) circle (2pt);
        \end{scope}
\end{tikzpicture}\label{4ptfunction}
\end{align}
where the configuration of topological lines is illustrated on the right-hand side.
This correlator contains information about the S-matrix element $S_{dc}^{ab}(\theta)$, which can be extracted using a modified LSZ reduction formula\footnote{The modified LSZ reduction formula refers to an extension of the standard LSZ formalism that can be applied to cases where the non-genuine local operators are concerned, as in our case.}.
The modified LSZ formula is derived in a similar manner to the conventional one (see, for example, Chapter 7 of~\cite{Peskin:1995ev}).

The most significant difference between the modified LSZ formula and its conventional counterpart lies in the normalization of the functions used during the amputation process.
To illustrate this, consider four spacetime points $y_1,y_2,z_1,z_2$. We examine the following correlation functions in the limit $|y_i-z_j|\to \infty$, which are relevant for amputating the initial external lines:
\begin{align}
    \braket{\phi_{dc}(y_1)\phi_{dc}(y_2) \phi_{cb}(z_1)\phi_{cb}(z_2)}&\ =\ 
    \begin{tikzpicture}[baseline=0cm]
   \begin{scope}[scale=0.63, shift={(0.5cm,0)}]
    \draw[thick, black!20!green]($(3.182,0)$) arc (0:45: 2cm);
    \draw[thick, black!20!green]($(3.182,0)$) arc (0:-45: 2cm);
    \draw[thick, black!20!green]($(4.828,0)$) arc (180:135: 2cm);
    \draw[thick, black!20!green]($(4.828,0)$) arc (180:225: 2cm);
    \draw [blue!80, thick] (4,0) circle (2);
    \node at ($(4.5,0) + (0:1)$) {$b$}; 
    \node at ($(4,1.2)$) {$c$}; 
    \node at ($(4,-1.2)$) {$c$}; 
    \node at ($(3.5,0) + (180:1)$) {$d$}; 
    \fill[black] (5.414, 1.414) circle (2pt);
    \fill[black] (5.414, -1.414) circle (2pt);
    \fill[black] (2.586, 1.414) circle (2pt);
    \fill[black] (2.586, -1.414) circle (2pt);
    \end{scope}
    \end{tikzpicture}\label{correctamputatefunction}\\
    \braket{\phi_{dc}(y_1)\phi_{dc}(y_2)}
    \braket{\phi_{cb}(z_1)\phi_{cb}(z_2)}
    &\ =\  \begin{tikzpicture}[baseline=0cm]
    \begin{scope}[scale=0.8, shift={(1.8cm, 0)}]
    \draw[thick, black!20!green] (8.5,-1) -- (8.5, 0);
    \draw[thick, black!20!green] (8.5, 0) -- (8.5, 1);
    \draw [blue!80, thick] (8.5,0) circle (1);
    \fill[black] (8.5, 1) circle (2pt);
    \fill[black] (8.5, -1) circle (2pt);
    \node at ($(8,0)$) {$d$}; 
    \node at ($(9,0)$) {$c$};
    \node at ($(11,0)$) {$c$};
    \draw[thick, black!20!green] (11.5,-1) -- (11.5, 0);
    \draw[thick, black!20!green] (11.5, 0) -- (11.5, 1);
    \draw [blue!80, thick] (11.5,0) circle (1);
    \fill[black] (11.5, 1) circle (2pt);
    \fill[black] (11.5, -1) circle (2pt);
    \node at ($(12,0)$) {$b$}; 
    \end{scope}
\end{tikzpicture}\label{wrongamputatefunction}
\end{align}
where the green lines indicate the contributions from poles corresponding to the kink states represented by those lines.
If the topological lines are invertible and anomaly-free, the correlation functions \eqref{correctamputatefunction} and \eqref{wrongamputatefunction} coincide.
However, if non-invertible lines are involved, they are not identical to each other in general. 
Indeed, we will shortly see that the following simple relation holds:
\begin{align}
    \begin{tikzpicture}[baseline=0cm]
   \begin{scope}[scale=0.63, shift={(0.5cm,0)}]
    \draw[thick, black!20!green]($(3.182,0)$) arc (0:45: 2cm);
    \draw[thick, black!20!green]($(3.182,0)$) arc (0:-45: 2cm);
    \draw[thick, black!20!green]($(4.828,0)$) arc (180:135: 2cm);
    \draw[thick, black!20!green]($(4.828,0)$) arc (180:225: 2cm);
    \draw [blue!80, thick] (4,0) circle (2);
    \node at ($(4.5,0) + (0:1)$) {$b$}; 
    \node at ($(4,1.2)$) {$c$}; 
    \node at ($(4,-1.2)$) {$c$}; 
    \node at ($(3.5,0) + (180:1)$) {$d$}; 
    \fill[black] (5.414, 1.414) circle (2pt);
    \fill[black] (5.414, -1.414) circle (2pt);
    \fill[black] (2.586, 1.414) circle (2pt);
    \fill[black] (2.586, -1.414) circle (2pt);
    \end{scope}
    \end{tikzpicture}
    \ = \ C_{dcb}\ \begin{tikzpicture}[baseline=0cm]
    \begin{scope}[scale=0.8, shift={(1.8cm, 0)}]
    \draw[thick, black!20!green] (8.5,-1) -- (8.5, 0);
    \draw[thick, black!20!green] (8.5, 0) -- (8.5, 1);
    \draw [blue!80, thick] (8.5,0) circle (1);
    \fill[black] (8.5, 1) circle (2pt);
    \fill[black] (8.5, -1) circle (2pt);
    \node at ($(8,0)$) {$d$}; 
    \node at ($(9,0)$) {$c$};
    \node at ($(11,0)$) {$c$};
    \draw[thick, black!20!green] (11.5,-1) -- (11.5, 0);
    \draw[thick, black!20!green] (11.5, 0) -- (11.5, 1);
    \draw [blue!80, thick] (11.5,0) circle (1);
    \fill[black] (11.5, 1) circle (2pt);
    \fill[black] (11.5, -1) circle (2pt);
    \node at ($(12,0)$) {$b$}; 
    \end{scope}
\end{tikzpicture}\label{Cdcb}
\end{align}
where $C_{dcb}$ is some constant depending on the topological lines.
It is important to emphasize that in extracting the S-matrix element $S^{ab}_{dc}(\theta)$ from the four-point function~\eqref{4ptfunction}, one must use the correlation function~\eqref{correctamputatefunction} rather than~\eqref{wrongamputatefunction}. 

We next see how the constant $C_{dcb}$ comes in the discussion of the crossing relation. For this purpose, let us introduce an auxiliary quantity $\Sw^{ab}_{dc}(\theta)$ obtained by incorrectly amputating the four-point function~\eqref{4ptfunction} using the correlator~\eqref{wrongamputatefunction}.
By construction, $\Sw^{ab}_{dc}(\theta)$ satisfies the ``crossing symmetry''
\begin{align}
    \Sw^{ab}_{dc}(\theta)=\Sw^{bc}_{ad}(i\pi - \theta)\ . \label{crossingSw}
\end{align}
The difference between $\Sw^{ab}_{dc}(\theta)$ and the correctly amputated S-matrix $S^{ab}_{dc}(\theta)$ lies solely in an overall normalization factor.
This constant factor can be read off from \eqref{Cdcb}, yielding the relation:
\begin{align}
    S^{ab}_{dc}(\theta)=\frac{1}{\sqrt{C_{dab}C_{dcb}}}\Sw^{ab}_{dc}(\theta)\ .
\end{align}
Combining this with \eqref{crossingSw}, we arrive at the modified crossing relation:
\begin{align}\label{eq:crossing}
        {S}^{ab}_{dc}(\theta)=\sqrt{\frac{{C}_{cba}\cdot {C}_{cda}}{{{C}_{dab}\cdot {C}_{dcb}}}}\, {S}^{da}_{cb}(\i\pi-\theta)\ .
\end{align}

Our remaining task is to derive the relation~\eqref{Cdcb} and evaluate the constant $C_{dcb}$. We begin by observing that the fusion property~\eqref{eq:fusionproperty} leads to the following decomposition:
\begin{align}
        \begin{tikzpicture}[baseline=0cm]
       \begin{scope}[scale=0.63, shift={(0.5cm,0)}]
        \draw[thick, black!20!green]($(3.182,0)$) arc (0:45: 2cm);
        \draw[thick, black!20!green]($(3.182,0)$) arc (0:-45: 2cm);
        \draw[thick, black!20!green]($(4.828,0)$) arc (180:135: 2cm);
        \draw[thick, black!20!green]($(4.828,0)$) arc (180:225: 2cm);
        \draw [blue!80, thick] (4,0) circle (2);
        \node at ($(4.5,0) + (0:1)$) {$b$}; 
        \node at ($(4,1.5)$) {$c$}; 
        \node at ($(4,-1.5)$) {$c$}; 
        \node at ($(3.5,0) + (180:1)$) {$d$}; 
        \fill[black] (5.414, 1.414) circle (2pt);
        \fill[black] (5.414, -1.414) circle (2pt);
        \fill[black] (2.586, 1.414) circle (2pt);
        \fill[black] (2.586, -1.414) circle (2pt);
        \end{scope}
        \node[right] at (4.5,0) {=\ \ \ {$\sum\limits_{e\in c \otimes c}\, \frac{{\d_{e}}}{\d_{c}}$}};
        \begin{scope}[scale=0.8, shift={(1.8cm, 0)}]
        \draw[thick, black!20!green] (8.5,-1) -- (8.5, 0);
        \draw[thick, black!20!green] (8.5, 0) -- (8.5, 1);
        \draw [blue!80, thick] (8.5,0) circle (1);
        \draw [blue!80, thick] (9.5,0)--(10.5, 0);
        \fill[blue!80] (9.5, 0) circle (2pt);
        \fill[blue!80] (10.5, 0) circle (2pt);
        \node[below] at (10, 0) {$e$};
        \fill[black] (8.5, 1) circle (2pt);
        \fill[black] (8.5, -1) circle (2pt);
        \node at ($(8,0)$) {$d$}; 
        \node at ($(9,0.5)$) {$c$};
        \node at ($(9,-0.5)$) {$c$};
        \node at ($(11,0.5)$) {$c$};
        \node at ($(11,-0.5)$) {$c$};
        \draw[thick, black!20!green] (11.5,-1) -- (11.5, 0);
        \draw[thick, black!20!green] (11.5, 0) -- (11.5, 1);
        \draw [blue!80, thick] (11.5,0) circle (1);
        \fill[black] (11.5, 1) circle (2pt);
        \fill[black] (11.5, -1) circle (2pt);
        \node at ($(12,0)$) {$b$}; 
        \end{scope}
\end{tikzpicture}
\end{align}
In the limit of large separation between the two bubbles, only the contribution from the identity line survives on the right-hand side\footnote{This can be understood as follows.
Let us consider the contribution from the topological line $\mathcal{L}_e$.
First, since the vacuum is an object in the regular module category, the action of $\mathcal{L}_e$ on the vacuum state yields $\ket{e}=\mathcal{L}_e \ket{\widetilde{0}}$.
Therefore, since this state is orthogonal to $\ket{\widetilde{0}}$ unless $e=1$, only the trivial line ($e=1$) contributes effectively.
The large separation limit is taken in order to suppress the overlapping between $\ket{e}$ and $\ket{\widetilde{0}}$.\label{footnote}}.  
Then, by comparing this with \eqref{Cdcb}, we can completely fix the constant $C_{dcb}$ as
\begin{align}\label{eq:constant}
    C_{dcb}=\frac{1}{\d_{c}}\ . 
\end{align}
By substituting this expression into \eqref{eq:crossing}, we conclude that the crossing relation satisfied by the correctly normalized S-matrix differs from the conventional one~\eqref{eq:crossing_relation_conventional}, and is instead modified by the TQFT dynamics as follows:
\begin{align}\label{eq:crossing_relation_new}
    {S}^{ab}_{dc}(\theta)=\sqrt{\frac{\d_{a} \d_{c}}{\d_{b} \d_{d}}}\, {S}^{bc}_{ad}(\i\pi-\theta)\ . 
\end{align}
This modified crossing relation is identical to to the one obtained in \cite{Copetti:2024rqj, Copetti:2024dcz}.

For later convenience, we end up this section by summarizing the discussions so far. When the theory possesses the integrability, the S-matrix ${S}^{ab}_{dc}(\theta)$ must be subject to the following three consistency conditions.
\begin{itemize}
    \item Unitarity condition:
    \begin{align}\label{eq:uni_new}
            \sum_{b}{S}^{bc}_{ad}(\theta)\, {S}^{ec}_{ab}(-\theta)=\delta_{d e}\ .
    \end{align}
    \item Yang-Baxter equation:
    \begin{align}\label{YBeq_new}
        \sum_{g}{S}^{gd}_{fe}(\theta_1 -\theta_2)\,{S}^{bc}_{gd}(\theta_1 -\theta_3)\,{S}^{ab}_{fg}(\theta_2 -\theta_3)=\sum_{g}{S}^{gc}_{ed}(\theta_2 -\theta_3)\,{S}^{ag}_{fe}(\theta_1 -\theta_3)\,{S}^{bc}_{ag}(\theta_1 -\theta_2)\ .
    \end{align}
    \item Modified crossing relation:
    \begin{align}\label{eq:crossing_relation_new2}
    {S}^{ab}_{dc}(\theta)=\sqrt{\frac{\d_{a} \d_{c}}{\d_{b} \d_{d}}}\, {S}^{bc}_{ad}(\i\pi-\theta)\ . 
\end{align}
\end{itemize}
\subsection{Gapped Theories from the Minimal Models}\label{subsec:IR_minimal}
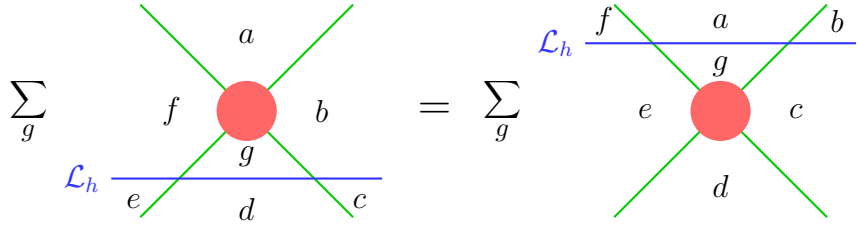
\begin{figure}
\centering
    \begin{tikzpicture}
       \node[right] at (-1.3, 0) {{\large $\sum\limits_{g}$}};
        \foreach \angle in {45, 135, 225, 315} {
            \draw[thick, black!20!green] (2,0) -- ++(\angle:2);
        }
        \fill [red!60] (2,0) circle (0.4);
        \draw[thick, blue!80] (0.2, -0.9) node[left]{$\mathcal{L}_{h}$}--(3.8, -0.9);
        \node at ($(2,0) + (0:1)$) {$b$}; 
        \node at ($(2,0) + (90:1)$) {$a$}; 
        \node at ($(2,0) + (180:1)$) {$f$}; 
        \node at ($(2,-0.3) + (270:1)$) {$d$};
        \node at ($(2,-0.6)$) {$g$};
        \node at ($(3.5,-0.2) + (270:1)$) {$c$};
        \node at ($(0.5,-0.2) + (270:1)$) {$e$};
        \node at (4.5, 0) {{\Large $=$}};
        \node[right] at (5, 0) {{\large $\sum\limits_{g}$}};
        \foreach \angle in {45, 135, 225, 315} {
            \draw[thick, black!20!green] (8.3,0) -- ++(\angle:2);
        }
        \fill [red!60] (8.3,0) circle (0.4);
        \draw[thick, blue!80] (6.5, 0.9) node[left]{$\mathcal{L}_{h}$}--(10.1, 0.9);
        \node at ($(8.3,1.2)$) {$a$};
        \node at ($(9.85,1.2)$) {$b$}; 
        \node at ($(9.3,0)$) {$c$};
        \node at ($(8.3,-1)$) {$d$};
        \node at ($(7.3,0)$) {$e$};
        \node at ($(6.75,1.2)$) {$f$}; 
        \node at ($(8.3,0.6)$) {$g$};
\end{tikzpicture}
\caption{Ward-Takahashi identities associated to the categorical symmetry $\mathcal{A}_p$.}\label{fig:WTid_bulk}
\end{figure}

Our prominent interest is the scattering theory in the gapped theories which are obtained via the relevant deformation of minimal models $M_{p}$ $(p \geq 4)$ by the primary operator $\Phi_{(1,3)}$:\footnote{The reason why we exclude the case $p=3$ is as follows. For this case, the corresponding UV CFT is the Ising model, and the $\Phi_{1,3}$-deformation preserves only $\mathbb{Z}_{2}$ spin-flip symmetry. Thereby, we cannot realize the scenario where non-invertible symmetries are spontaneously broken.}
\begin{align}
    S_{p}[\lambda]=S_{p}+\lambda\int \text{d} t\,\text{d} x \, \Phi_{(1,3)}(t,x)\ . 
\end{align}
Here, $S_{p}$ is the CFT action of the minimal model $M_{p}$, and $S_{p}[\lambda]$ is the deformed action. Also, $\lambda$ is the relevant coupling, and $\Phi_{(1,3)}$ is the primary operator in $M_{p}$ with Kac label~$(1,3)$. There are two important remarks in order. Firstly, this deformation does not break the integrability i.e.,  there are infinitely many higher-spin currents conserved along with the RG flow \cite{Zamolodchikov:1989hfa, Zamolodchikov:1991vx}. Secondly, this relevant deformation preserves the categorical symmetry $\mathcal{A}_{p}$ whose fusion rule is given by \eqref{fusion_rule}.
This fusion category $\mathcal{A}_p$ is spontaneously broken, and we have $(p-1)$-fold degenerate vacua $\{\ket{a}\}$ defined by \eqref{eq:vacuum_states}. Accordingly, we can introduce $(p-2)$ kinks interpolating adjacent vacua $\ket{a}$ and $\ket{a+1}$, which allows us to consider the kink-scattering amplitudes.
By solving the three conditions from \eqref{eq:uni_new} to \eqref{eq:crossing_relation_new2}, one can derive the correctly normalized S-matrix given by \cite{Copetti:2024rqj}:\footnote{It is worth mentioning the comparison to the old literature. The correctly normalized S-matrix ${S}^{ab}_{dc}$ in \eqref{eq:S_matrix_RSOS_new} differs from the traditionally used one $\widehat{S}^{ab}_{dc}$ in the old literature by some factor:
\begin{align}\label{eq:S_matrix_comparison}
   {S}^{ab}_{dc}(\theta)=\left(\frac{\d_{b} \d_{d}}{\d_{a}\d_{c}}\right)^{\frac{\i\theta}{2\pi}}\widehat{S}^{ab}_{dc}(\theta)\ . 
\end{align}}
\begin{align}\label{eq:S_matrix_RSOS_new}
    {S}^{ab}_{dc}(\theta)=Z_{p}(\theta)\left[\delta_{bd}\left(\frac{\d_{a}\,\d_{c}}{\d_{b}\,\d_{d}}\right)^{\frac{1}{2}}\sinh\left(\frac{\theta}{p}\right)+\delta_{ac}\sinh\left(\frac{\i\pi-\theta}{p}\right)\right]\ , 
\end{align}
where $\d_{a}$ is the quantum dimension of the topological defect line $\mathcal{L}_{a}$:
\begin{align}\label{eq:quantum_dimension}
    \d_{a}=\frac{\sin\left(a\pi/p\right)}{\sin\left(\pi/p\right)}\ ,
\end{align}
and the overall factor $Z_{p}(\theta)$ is given by 
\begin{align}\label{eq:explicit_form_Z}
    \begin{aligned}
        Z_{p}(\theta)&=\frac{1}{\i\pi}\Gamma\left(\frac{1}{p}\right)\Gamma\left(1+\frac{\i\theta}{p\pi}\right)\Gamma\left(1-\frac{1}{p}-\frac{\i\theta}{p\pi}\right)\prod_{\ell=1}^{\infty}\frac{F_{\ell, p}(\theta)F_{\ell, p}(\i\pi-\theta)}{F_{\ell, p}(0)F_{\ell, p}(\i\pi)}\ , \\
        F_{\ell, p}(\theta)&:=\frac{\Gamma\left(\frac{2\ell}{p}+\frac{\i\theta}{p \pi}\right)\Gamma\left(1+\frac{2\ell}{p}+\frac{\i\theta}{p \pi}\right)}{\Gamma\left(\frac{2\ell+1}{p}+\frac{\i\theta}{p \pi}\right)\Gamma\left(1+\frac{2\ell-1}{p}+\frac{\i\theta}{p \pi}\right)}\ . 
    \end{aligned}
\end{align}
In particular, the factor $Z_{p}(\theta)$ satisfies the following conditions:
\begin{align}
    Z_{p}(\theta)Z_{p}(-\theta)&=\frac{1}{\sinh\left(\frac{\i\pi-\theta}{p}\right)\sinh\left(\frac{\i\pi+\theta}{p}\right)}\ , \\
    Z_{p}(\i\pi-\theta)&=Z_{p}(\theta)\ .
\end{align}
We can check that this S-matrix ${S}^{ab}_{dc}(\theta)$ satisfies the Ward-Takahashi identities associated to the non-invertible symmetries as described in Fig.\,\ref{fig:WTid_bulk}.

\section{Boundary Scattering and Non-invertible Symmetries}\label{sec:boundary_scattering}
So far, we have reviewed the scattering theories in $1+1$ dimensions, and the correctly normalized S-matrix respecting the non-invertible symmetries. The main goal of this section is to extend these discussions to the QFTs in the presence of boundaries. Under the assumption that the boundary is weakly-symmetric for some of the non-invertible symmetries in the UV theory, we will see that the boundary crossing relation should be also modified.
\subsection{Boundary S-matrix}\label{subsec:boundary_S_matrix}
We first introduce key concepts of the boundary scattering following \cite{Ghoshal:1993tm}.
With a spacetime boundary, we can define boundary scattering alongside bulk scattering. Throughout this section, we adopt the same assumptions as in the previous section since boundaries generally do not alter bulk dynamics.  We place a boundary $\mathcal{B}$ along the time direction at $x=0$, and consider the scattering in the left half plane. A kink $K_{ab}$ created at some bulk point far from the boundary can scatter off the boundary, producing a kink $K_{ac}$. The boundary S-matrix~$R^{a}_{\,bc}(\theta)$ is then defined by the following relation:
\begin{align}\label{eq:def_R_matrix}
    \ket{K_{ab}(\theta)}_{\mathcal{B}, \text{in}}=\sum_{c}R^{a}_{\,bc}(\theta)\ket{K_{ac}(-\theta)}_{\mathcal{B}, \text{out}}+\cdots\ , 
\end{align}
where $\ket{K_{ab}(\theta)}_{\mathcal{B}, \text{in/out}}$ is the asymptotic state belonging to the Hilbert space $\mathcal{H}_{\mathcal{B}}$ associated to an infinite half line $x\in (-\infty, 0]$ with constant time. 
In a similar way to the bulk S-matrix~\eqref{eq:graphical_S_matrix}, we express the boundary S matrix as follows:
\begin{align}\label{eq:R-matrix}
    \begin{tikzpicture}[baseline=0cm]
       \node[right] at (0,0) {$R^{a}_{\, bc}(\theta)\ \ =$}; 
        \foreach \angle in {225} {
             \draw[thick, black!20!green] (4,0) -- ++(\angle:1);
         }
         \draw[thick, black!20!green, ->] ($(4,0) + (225:2)$) -- ++(45:1);
        \foreach \angle in {135} {
             \draw[thick, black!20!green, ->] (4,0) -- ++(\angle:1.3);
             \draw[thick, black!20!green] ($(4,0) + (\angle:1)$) -- ++(\angle:1);
         }
        \node [black!20!green, below left] at ($(4,0) + (225:1.8)$) {$\theta$};
        \fill [black!20!yellow] (4,0) circle (0.4);
        \node at ($(3.7,1)$) {$c$}; 
        \node at ($(4,0) + (180:1)$) {$a$}; 
        \node at ($(3.7,-1)$) {$b$};
        \fill[black!20!white] (4, 1.614)--(4.5, 1.614)--(4.5, -1.614)--(4, -1.614);
        \draw[purple, ultra thick] (4, 1.614)--(4, -1.614);
\end{tikzpicture}
\end{align}
In general, the presence of a boundary in spacetime breaks both the fusion categorical symmetry $\mathcal{A}_{p}$ and integrability of the bulk theory. However, in this paper, we focus on boundaries that preserve these structures and discuss the corresponding scattering processes. Just as the bulk S-matrix is constrained by several conditions discussed in Sec.~\ref{subsec:modified_crossing_bulk}, the boundary S-matrix must also satisfy certain consistency requirements. These include the unitarity, boundary Yang-Baxter equations, and boundary crossing relation, which we will describe in order.
\subsubsection*{Boundary unitarity condition}
Similar to the bulk S-matrix, the boundary S-matrix should be a unitary transformation acting on the asymptotic kink states as defined in \eqref{eq:def_R_matrix}. Therefore, when the boundary integrability is preserved\footnote{If boundary integrability is absent, multi-kink solitons can contribute as intermediate processes.}, the boundary S-matrix must satisfy the following unitarity condition:
\begin{align}\label{eq:bdy_unitary}
    \sum_{d}R^{b}_{\, cd}(\theta)R^{b}_{\, da}(-\theta)=\delta_{ac}\ , 
\end{align}
\begin{align}\label{eq:graphical_unitarity_boundary}
    \begin{tikzpicture}[scale=0.8, baseline=0.7cm]
        \node[right] at (0.5,0.8) {{\Large $\sum\limits_{d}$}};
        \foreach \angle in {225} {
             \draw[thick, black!20!green] (4,0) -- ++(\angle:1);
         };
         \draw[thick, black!20!green, ->] ($(4,0) + (225:1.5)$) -- ++(45:0.75);
         \draw[thick, black!20!green, ->] ($(4,0) + (315:1.5)$) -- ++(135:0.75);
        \foreach \angle in {135} {
             \draw[thick, black!20!green, ->] (4,2) -- ++(\angle:1);
             \draw[thick, black!20!green] ($(4,2) + (\angle:0.75)$) -- ++(\angle:0.75);
         };
        \draw[thick, black!20!green][<->] ($(4,1) + (0:0.8)$) arc (0:-180: 0.8cm and 1cm);
        \draw[thick, black!20!green][] ($(4,1) + (0:0.8)$) arc (0:180: 0.8cm and 1cm);
        \node [black!20!green] at ($(4,0) + (225:1.8)$) {$\theta$};
        \fill [black!20!yellow] (4,0) circle (0.4);
        \fill [black!20!yellow] (4,2) circle (0.4);
        \node at ($(4,1) + (180:1.5)$) {$b$}; 
        \node at ($(3.6,0) + (270:1)$) {$c$};
        \node at ($(3.65,0) + (90:1)$) {$d$};  
        \node at ($(3.6,2) + (90:1)$) {$a$};
        \node at (7,1) {{\Large =}};
        \draw[thick, black!20!green, ->] (10,-1.3) -- (10, 0.9);
        \draw[thick, black!20!green] (10, 0.9) -- (10, 3.3);
        \node at ($(10.75,1)$) {$c$}; 
        \node at ($(9, 1)$) {$b$}; 
        \node at (8,1) {{\large $\delta_{ac}$}};
        \fill[black!20!white] (4, 3.3)--(5.5, 3.3)--(5.5, -1.3)--(4, -1.3);
        \draw[purple, ultra thick] (4, 3.3)--(4, -1.3);
        \fill[black!20!white] (11.5, 3.3)--(13, 3.3)--(13, -1.3)--(11.5, -1.3);
        \draw[purple, ultra thick] (11.5, 3.3)--(11.5, -1.3);
\end{tikzpicture}
\end{align}
\subsubsection*{Boundary Yang-Baxter equation}
As an analog of the Yang-Baxter equation \eqref{eq:YBeq},  when both of the bulk and boundary integrability exist, we have the boundary Yang-Baxter equation:
\begin{align}\label{eq:bdy_YB}
    \begin{aligned}
        &\sum_{f, g}S^{g a}_{c b}(\theta_1 -\theta_2)R^{g}_{\,af}(\theta_1 )S^{d f}_{c g}(\theta_1 +\theta_2)R^{d}_{\,fe}(\theta_2 ) \\
        &\qquad \qquad =\sum_{f, g} R^{b}_{\, ag}(\theta_2 ) S^{f g}_{c b}(\theta_1 +\theta_2 )R^{f}_{\, ge}(\theta_1 ) S^{de}_{c f}(\theta_1 -\theta_2 )\ ,
    \end{aligned}
\end{align}
\begin{align}
 \begin{tikzpicture}[baseline=0cm] 
        \node[right] at (-0.5,0) {{\Large $\sum\limits_{f, g}$}};
         \draw[thick, black!20!green] (2, -2) -- ++(45:{2*sqrt(2)})--(2, 2);
         \node[black!20!green, below right] at (2, -2) {$\theta_2$};
         \draw[thick, black!20!green] (1, -2) -- ++(20:{3/cos(20)})--++(160:{3/cos(20)});
         \node[black!20!green, below left] at (1, -2) {$\theta_1$};
         \fill[black!20!white] (4, 2.5)--(5.2, 2.5)--(5.2, -2.5)--(4, -2.5);
        \draw[purple, ultra thick] (4, -2.5)--(4, 2.5);
        \node at (3, -2) {$a$};
        \node at (1.7, -2) {$b$};
        \node at ($(-1.5,-2)+(20:{3/cos(20)})$) {$c$};
        \node at (2.35, 0.5) {$d$};
        \node at (3.6, 1) {$e$};
        \node at (3.8, -0.5) {$f$};
        \node at (3.34, -0.9) {$g$};
         \node[right] at (6,0) {{\Large $=$}};
         \node[right] at (7.5,0) {{\Large $\sum\limits_{f, g}$}};
         \draw[thick, black!20!green] (10, -2) -- ++(45:{2*sqrt(2)})--(10, 2);
         \node[black!20!green, below] at (10, -2) {$\theta_2$};
         \draw[thick, black!20!green] (9, 2) -- ++(-20:{3/cos(20)})--++(-160:{3/cos(20)}) node[black!20!green, below]{$\theta_1$};
         \fill[black!20!white] (12, 2.5)--(13.2, 2.5)--(13.2, -2.5)--(12, -2.5);
        \draw[purple, ultra thick] (12, -2.5)--(12, 2.5);
        \node at (11.5, -1.3) {$a$};
        \node at (10.2, -0.8) {$b$};
        \node at (9.7, 1) {$c$};
        \node at (9.7, 2) {$d$};
        \node at (11.3, 1.7) {$e$};
        \node at (11.4, 0.9) {{\footnotesize $f$}};
        \node at (11.8, 0.5) {$g$};
\end{tikzpicture}
\end{align}
\subsubsection*{Boundary crossing relation}
To give more constraints on the boundary S-matrix, we need the boundary analog of the crossing relation. However, unlike the previous two conditions, it is not immediately obvious. To gain insight into this, we need to shift our perspective: instead of considering a boundary extending in the time direction as before, we now transition to a picture where the boundary extends in the spatial direction. Under this reinterpretation, one can define a new matrix $\mathsf{R}^{a}_{\, bc}(\theta)$ associated with the boundary scattering process by interchanging the roles of time and space in the original boundary S-matrix \eqref{eq:R-matrix}:
\begin{align}
            \begin{tikzpicture}[baseline=0cm]
                   \node[right] at (-8,0) {$R^{a}_{\, bc}(\theta)\ \ =$}; 
        \foreach \angle in {225} {
             \draw[thick, black!20!green] (-4,0) -- ++(\angle:1);
         }
         \draw[thick, black!20!green, ->] ($(-4,0) + (225:2)$) -- ++(45:1);
        \foreach \angle in {135} {
             \draw[thick, black!20!green, ->] (-4,0) -- ++(\angle:1.3);
             \draw[thick, black!20!green] ($(-4,0) + (\angle:1)$) -- ++(\angle:1);
         }
        \node [black!20!green, below left] at ($(-4,0) + (225:1.8)$) {$\theta$};
        \fill [black!20!yellow] (-4,0) circle (0.4);
        \node at ($(3.7-8,1)$) {$c$}; 
        \node at ($(4-8,0) + (180:1)$) {$a$}; 
        \node at ($(3.7-8,-1)$) {$b$};
        \fill[black!20!white] ($(4-8, 1.614)$)--($(4.5-8, 1.614)$)--($(4.5-8, -1.614)$)--($(4-8, -1.614)$);
        \draw[purple, ultra thick] (-4, 1.614)--(-4, -1.614);
       \node[right] at (-0.3,0) {$\mathsf{R}^{a}_{\,bc}(\theta)\ \ =$}; 
            \draw[thick, black!20!green, ->] (4,-0.5) -- ++(45:1.2);
            \draw[thick, black!20!green] ($(4,-0.5)+(45:1.2)$) -- ++(45:0.7);
            \draw[thick, black!20!green] (4,-0.5) -- ++(135:1.1);
            \draw[thick, black!20!green, <-] ($(4,-0.5)+(135:1.1)$) -- ++(135:0.8);
        \node [black!20!green] at ($(3.6,2) + (225:1.8)$) {$\theta$};
        \fill[black!20!yellow] (4.4,-0.5) arc[start angle=0, end angle=180, radius=0.4cm] -- cycle;
        \node at ($(4,-0.5) + (20:1)$) {$c$}; 
        \node at ($(4,-0.5) + (90:1)$) {$a$}; 
        \node at ($(4,-0.5) + (160:1)$) {$b$};
        \fill[black!20!white] (2.5, -0.5)--(5.5, -0.5)--(5.5, -1)--(2.5, -1)--cycle;
        \draw[purple, ultra thick] (2.5,-0.5)--(5.5,-0.5);
        \node[right] at (-2.5,0) {{\Large $\longrightarrow$}};
\end{tikzpicture}
\end{align}
Physically, this new matrix $\mathsf{R}$ describes the probability of a kink pair production from the boundary state $\ket{\mathcal{B}}$:
\begin{align}
    \ket{\mathcal{B}}=\sum_{a,b,c}\mathsf{R}^{a}_{\,bc}\left(\frac{\i\pi}{2}-\theta\right)\ket{K_{ba}(-\theta)K_{ac}(\theta)}_{\text{out}}+\dots.
\end{align}
We should make important comments here. In the rotated picture, the matrix $\mathsf{R}$ is the map between {\it bulk} Hilbert spaces since the time evolves in the transverse direction to the boundary. This is a striking difference from the boundary S-matrix $R$ in the pre-rotated picture, where the Hilbert space is associated to an infinite {\it half} line. Since emitted kinks from the boundary belong to the bulk Hilbert space, in and out kink asymptotic states must be related by the bulk S-matrix. Based on this observation, 
Ghoshal and Zamolodchikov proposed the following boundary crossing relation~\cite{Ghoshal:1993tm}:
\begin{align}\label{eq:bdy_crossing_relation}
    \mathsf{R}^{a}_{\, bc}(\theta)&\overset{?}{=}R^{a}_{\, bc}(\theta)\ , \\
    \mathsf{R}^{b}_{\, ca}\left(\frac{\i\pi}{2}-\theta\right)&=\sum_{d}S^{ba}_{cd}(2\theta)\, \mathsf{R}^{d}_{\, ca}\left(\frac{\i\pi}{2}+\theta\right)\ .
\end{align}
Remark that the second equation holds only when bulk integrability is present.
It is convenient to express the second equation in a diagrammatic manner:
\begin{align}\label{eq:crossing_bdy}
    \begin{tikzpicture}[baseline=0cm]
    \draw[thick, black!20!green, ->] (-4,-0.5) -- ++(45:1.2);
            \draw[thick, black!20!green] ($(-4,-0.5)+(45:1.2)$) -- ++(45:0.7);
            \draw[thick, black!20!green] (-4,-0.5) -- ++(135:1.1);
            \draw[thick, black!20!green, <-] ($(-4,-0.5)+(135:1.1)$) -- ++(135:0.8);
        \fill[black!20!yellow] (-3.6,-0.5) arc[start angle=0, end angle=180, radius=0.4cm] -- cycle;
        \node at ($(-4,-0.5) + (20:1)$) {$a$}; 
        \node at ($(-4,-0.5) + (90:1)$) {$b$}; 
        \node at ($(-4,-0.5) + (160:1)$) {$c$};
        \fill[black!20!white] (-5.5, -0.5)--(-2.5, -0.5)--(-2.5, -1)--(-5.5, -1)--cycle;
        \draw[purple, ultra thick] (-5.5,-0.5)--(-2.5,-0.5);
        \node[right] at (-1,0.2) {{\Large $=$}};
        \node[right] at (0.5,0.3) {{\Large $\sum\limits_{d}$}};
         \draw[thick, black!20!green, ->] ($(4,1.5) + ({1.5/sqrt(2)}, {1.5/sqrt(2)})$) -- ++(225:0.7);
         \draw[thick, black!20!green] ($(4,1.5) + ({1.5/sqrt(2)}, {1.5/sqrt(2)}) + (225:0.7)$) -- ++(225:0.8);
         \draw[thick, black!20!green, ->] (4,1.5) -- ++(135:1);
         \draw[thick, black!20!green] ($(4,1.5) + (135:0.75)$) -- ++(135:0.75);
        \draw[thick, black!20!green][<-] ($(4,0.5) + (0:0.8)$) arc (0:-180: 0.8cm and 1cm);
        \draw[thick, black!20!green][->] ($(4,0.5) + (0:0.8)$) arc (0:180: 0.8cm and 1cm);
        \fill [black!20!yellow] (4,-0.5) circle (0.4);
        \fill [red!60] (4,1.5) circle (0.4);
        \node at ($(4,0.5) + (0:1.5)$) {$a$}; 
        \node at ($(4,-0.5) + (90:1)$) {$d$}; 
        \node at ($(4,0.5) + (180:1.5)$) {$c$}; 
        \node at ($(4,1.5) + (90:1)$) {$b$};
        \fill[black!20!white] (5.5, -0.5)--(2.5, -0.5)--(2.5, -1)--(5.5, -1)--cycle;
        \draw[purple, ultra thick] (5.5,-0.5)--(2.5,-0.5);
\end{tikzpicture}
\end{align}
In numerous studies including \cite{Chim:1995kf, Ahn:1996nq, DeMartino:1998br, Nepomechie:2002tw, Bajnok:2002vp, Dorey:2005ak, Kruczenski:2020ujw}, the calculation of the boundary S-matrix has been carried out by solving the three consistency conditions discussed above. However, as we will see in the next section, the boundary crossing relation is also modified, just as the bulk crossing relation is done.
 
\subsection{Modified Boundary Crossing Relation}
In this section, we discuss the correct normalization of the boundary S-matrix by carefully keeping track of the TQFT effects coming from the spontaneously breaking of non-invertible symmetries. We start with the analysis on the boundary S-matrix $R^{a}_{\, bc}$. To do this, we first consider the two-point correlation function in the presence of the boundary:
\begin{align}\label{eq:two_pt}
        \begin{tikzpicture}[baseline=0cm]
        \pgfmathsetmacro{\radius}{1.5/sqrt(2)}
        \node[right] at (-11.5,0) {$\langle \phi_{ab}(x_{1})\phi_{ac}(x_{2})\rangle_{\mathcal{B}}\ =\ $}; 
        \begin{scope}[shift={(-1.4cm, 0)}]
        \node at ($(3.7-8-0.2,1+0.2)$) {$c$}; 
        \node at ($(4-8-0.3,0) + (180:1)$) {$a$}; 
        \node at ($(3.7-8-0.2,-1-0.2)$) {$b$};
        \fill[black!20!white] ($(4-8, 1.914)$)--($(4.5-8, 1.914)$)--($(4.5-8, -1.914)$)--($(4-8, -1.914)$);
        \draw[thick, blue!80] ($(-4, 1.5)$) arc (90:270: 1.5cm);
                 \foreach \angle in {135, 225} {
             \fill[black] ($(-4,0)+(\angle:1.5)$) circle(2pt);
        \draw[purple, ultra thick] (-4, 1.914)--(-4, -1.914);
         }
        \end{scope}
\end{tikzpicture}
\end{align}
where the vacuum is given by $\ket{\widetilde{0}}_{\mathcal{B}}$. The reason why we attach the subscript $\mathcal{B}$ is to emphasize that this vacuum state belongs to the boundary Hilbert space $\mathcal{H}_{\mathcal{B}}$ rather than the bulk one. By making use of the modified LSZ reduction formula, one can extract the boundary S-matrix $R^{a}_{\, bc}(\theta)$ from this correlator. As in the bulk case, we must carefully keep track with the normalization which is relevant for the amputation. Consider two bulk spacetime points
$y$ and $z$ such that they are located far away from the boundary $|y_{\perp}|\, ,  |z_{\perp}|\, \to\infty$.  Here, $y_{\perp}$ and $z_{\perp}$ are the spacetime coordinates which are transverse to the boundary. Then, the modified LSZ reduction formula concerns the following correlation function for the amputation of initial external lines:
\begin{align}\label{eq:correct_amputate}
        \begin{tikzpicture}[baseline=0cm]
        \pgfmathsetmacro{\radius}{1.5/sqrt(2)}
        \node[right] at (-12.5,0) {$\langle \phi_{ab}(y)\phi_{ab}(z)\rangle_{\mathcal{B}}\ =\ $}; 
        \begin{scope}[shift={(-2.4cm, 0)}]
        \node at ($(3.7-8-0.2,1+0.2)$) {$b$}; 
        \node at ($(4-8-0.3,0) + (180:1)$) {$a$}; 
        \node at ($(3.7-8-0.2,-1-0.2)$) {$b$};
        \fill[black!20!white] ($(4-8, 1.914)$)--($(4.5-8, 1.914)$)--($(4.5-8, -1.914)$)--($(4-8, -1.914)$);
        \draw[thick, blue!80] ($(-4, 1.5)$) arc (90:270: 1.5cm);
        \draw[thick, black!20!green] ($(-4,0)+(225:1.5)$) arc (-90:90: 0.5cm and \radius cm);
                 \foreach \angle in {135, 225} {
             \fill[black] ($(-4,0)+(\angle:1.5)$) circle(2pt);
        \draw[purple, ultra thick] (-4, 1.914)--(-4, -1.914);
         }
        \end{scope}
\end{tikzpicture}
\end{align}
while the conventional LSZ formula concerns
\begin{align}\label{eq:wrong_amputate}
        \begin{tikzpicture}[baseline=0cm]
        \pgfmathsetmacro{\radius}{1.5/sqrt(2)}
        \node[right] at (0,0) {$\langle \phi_{ab}(y_{1})\phi_{ab}(z_{1})\rangle\ =\ $}; 
        \begin{scope}[scale=0.8, shift={(-2.4cm, 0)}]
        \draw[thick, black!20!green] (8.5,-1) -- (8.5, 0);
        \draw[thick, black!20!green] (8.5, 0) -- (8.5, 1);
        \draw [blue!80, thick] (8.5,0) circle (1);
        \fill[black] (8.5, 1) circle (2pt);
        \fill[black] (8.5, -1) circle (2pt);
        \node at ($(8,0)$) {$a$}; 
        \node at ($(9,0)$) {$b$};
        \end{scope}
        \end{tikzpicture}
\end{align}
Likewise in the bulk case, when topological lines are non-invertible, these two correlation functions differ by some constant $C^{\mathcal{B}}_{ab}$ as
\begin{align}\label{eq:constCb}
    \begin{tikzpicture}[baseline=0cm]
        \pgfmathsetmacro{\radius}{1.5/sqrt(2)}
        \begin{scope}[scale=0.8, shift={(5.4cm, 0)}]
        \node at ($(3.7-8-0.2,1+0.2)$) {$b$}; 
        \node at ($(4-8-0.3,0) + (180:1)$) {$a$}; 
        \node at ($(3.7-8-0.2,-1-0.2)$) {$b$};
        \fill[black!20!white] ($(4-8, 1.914)$)--($(4.5-8, 1.914)$)--($(4.5-8, -1.914)$)--($(4-8, -1.914)$);
        \draw[thick, blue!80] ($(-4, 1.5)$) arc (90:270: 1.5cm);
        \draw[thick, black!20!green] ($(-4,0)+(225:1.5)$) arc (-90:90: 0.5cm and \radius cm);
                 \foreach \angle in {135, 225} {
             \fill[black] ($(-4,0)+(\angle:1.5)$) circle(2pt);
        \draw[purple, ultra thick] (-4, 1.914)--(-4, -1.914);
         }
        \end{scope}
        \begin{scope}[scale=0.8, shift={(-2.4cm, 0)}]
        \draw[thick, black!20!green] (8.5,-1) -- (8.5, 0);
        \draw[thick, black!20!green] (8.5, 0) -- (8.5, 1);
        \draw [blue!80, thick] (8.5,0) circle (1);
        \fill[black] (8.5, 1) circle (2pt);
        \fill[black] (8.5, -1) circle (2pt);
        \node at ($(8,0)$) {$a$}; 
        \node at ($(9,0)$) {$b$};
        \end{scope}
        \node[right] at (2,0) {$\ =\ C^{\mathcal{B}}_{ab}$}; 
        \end{tikzpicture}
\end{align}

Before discussing the relation between this constant $C^{\mathcal{B}}_{ab}$ with topological lines, we elaborate on the boundary S-matrix in the rotated picture. The boundary S-matrix $\mathsf{R}^{a}_{\, bc}$ can be pulled out from the following two-point function via the modified LSZ reduction formula:
\begin{align}\label{eq:two_pt_rotated}
    \begin{tikzpicture}[baseline=0cm]
    \node[right] at (-7,0) {$\langle \phi_{ab}(x_{1})\phi_{ac}(x_{2})\rangle_{\mathcal{B}}=\bra{\widetilde{0}} \text{T} \phi_{ba}(x_{1})\phi_{ac}(x_{2})\ket{\mathcal{B}} \ =\ $}; 
        \begin{scope}[shift={(4cm, -4.5cm)}, rotate=-90]
        \node at ($(3.7-8-0.2,1+0.2)$) {$c$}; 
        \node at ($(4-8-0.3,0) + (180:1)$) {$a$}; 
        \node at ($(3.7-8-0.2,-1-0.2)$) {$b$};
        \fill[black!20!white] ($(4-8, 1.914)$)--($(4.5-8, 1.914)$)--($(4.5-8, -1.914)$)--($(4-8, -1.914)$);
        \draw[thick, blue!80] ($(-4, 1.5)$) arc (90:270: 1.5cm);
                 \foreach \angle in {135, 225} {
             \fill[black] ($(-4,0)+(\angle:1.5)$) circle(2pt);
        \draw[purple, ultra thick] (-4, 1.914)--(-4, -1.914);
         }
        \end{scope}
    \end{tikzpicture}
\end{align}
where $\ket{\mathcal{B}}$ is the boundary state belonging to the bulk Hilbert space. We again encounter the time to discuss the normalization which is closely related to the amputation. Consider the four bulk spacetime points $y_1$, $y_{2}$, $z_{1}$ and $z_{2}$ such that they are far enough away i.e., $|y_{i}-z_{j}|\to\infty$. Then, the modified LSZ reduction formula concerns the following two-point function for amputating the out-going external lines:
\begin{align}
    \braket{\phi_{ba}(y_1)\phi_{ba}(y_2) \phi_{ac}(z_1)\phi_{ac}(z_2)}&\ =\ 
    \begin{tikzpicture}[baseline=0cm]
   \begin{scope}[scale=0.63, shift={(0.5cm,0)}]
    \draw[thick, black!20!green]($(3.182,0)$) arc (0:45: 2cm);
    \draw[thick, black!20!green]($(3.182,0)$) arc (0:-45: 2cm);
    \draw[thick, black!20!green]($(4.828,0)$) arc (180:135: 2cm);
    \draw[thick, black!20!green]($(4.828,0)$) arc (180:225: 2cm);
    \draw [blue!80, thick] (4,0) circle (2);
    \node at ($(4.5,0) + (0:1)$) {$c$}; 
    \node at ($(4,1.2)$) {$a$}; 
    \node at ($(4,-1.2)$) {$a$}; 
    \node at ($(3.5,0) + (180:1)$) {$b$}; 
    \fill[black] (5.414, 1.414) circle (2pt);
    \fill[black] (5.414, -1.414) circle (2pt);
    \fill[black] (2.586, 1.414) circle (2pt);
    \fill[black] (2.586, -1.414) circle (2pt);
    \end{scope}
    \end{tikzpicture}\label{correctamputatefunction_bdy}
\end{align}
while the conventional LSZ reduction formula concerns
\begin{align}
    \braket{\phi_{ba}(y_1)\phi_{ba}(y_2)}
    \braket{\phi_{ac}(z_1)\phi_{ac}(z_2)}
    &\ =\  \begin{tikzpicture}[baseline=0cm]
    \begin{scope}[scale=0.8, shift={(1.8cm, 0)}]
    \draw[thick, black!20!green] (8.5,-1) -- (8.5, 0);
    \draw[thick, black!20!green] (8.5, 0) -- (8.5, 1);
    \draw [blue!80, thick] (8.5,0) circle (1);
    \fill[black] (8.5, 1) circle (2pt);
    \fill[black] (8.5, -1) circle (2pt);
    \node at ($(8,0)$) {$b$}; 
    \node at ($(9,0)$) {$a$};
    \node at ($(11,0)$) {$a$};
    \draw[thick, black!20!green] (11.5,-1) -- (11.5, 0);
    \draw[thick, black!20!green] (11.5, 0) -- (11.5, 1);
    \draw [blue!80, thick] (11.5,0) circle (1);
    \fill[black] (11.5, 1) circle (2pt);
    \fill[black] (11.5, -1) circle (2pt);
    \node at ($(12,0)$) {$c$};
    \end{scope}
\end{tikzpicture}\label{wrongamputatefunction_bdy}
\end{align}
Fortunately, we have already seen the difference between these two correlators as \eqref{Cdcb}. For convenience, let us repeat it here
\begin{align}
    \begin{tikzpicture}[baseline=0cm]
   \begin{scope}[scale=0.63, shift={(0.5cm,0)}]
    \draw[thick, black!20!green]($(3.182,0)$) arc (0:45: 2cm);
    \draw[thick, black!20!green]($(3.182,0)$) arc (0:-45: 2cm);
    \draw[thick, black!20!green]($(4.828,0)$) arc (180:135: 2cm);
    \draw[thick, black!20!green]($(4.828,0)$) arc (180:225: 2cm);
    \draw [blue!80, thick] (4,0) circle (2);
    \node at ($(4.5,0) + (0:1)$) {$c$}; 
    \node at ($(4,1.2)$) {$a$}; 
    \node at ($(4,-1.2)$) {$a$}; 
    \node at ($(3.5,0) + (180:1)$) {$b$}; 
    \fill[black] (5.414, 1.414) circle (2pt);
    \fill[black] (5.414, -1.414) circle (2pt);
    \fill[black] (2.586, 1.414) circle (2pt);
    \fill[black] (2.586, -1.414) circle (2pt);
    \end{scope}
    \end{tikzpicture}
    \ = \ C_{bac}\ \begin{tikzpicture}[baseline=0cm]
    \begin{scope}[scale=0.8, shift={(1.8cm, 0)}]
    \draw[thick, black!20!green] (8.5,-1) -- (8.5, 0);
    \draw[thick, black!20!green] (8.5, 0) -- (8.5, 1);
    \draw [blue!80, thick] (8.5,0) circle (1);
    \fill[black] (8.5, 1) circle (2pt);
    \fill[black] (8.5, -1) circle (2pt);
    \node at ($(8,0)$) {$b$}; 
    \node at ($(9,0)$) {$a$};
    \node at ($(11,0)$) {$a$};
    \draw[thick, black!20!green] (11.5,-1) -- (11.5, 0);
    \draw[thick, black!20!green] (11.5, 0) -- (11.5, 1);
    \draw [blue!80, thick] (11.5,0) circle (1);
    \fill[black] (11.5, 1) circle (2pt);
    \fill[black] (11.5, -1) circle (2pt);
    \node at ($(12,0)$) {$c$}; 
    \end{scope}
\end{tikzpicture}\label{Cbac}
\end{align}
where $C_{bac}$ is given by \eqref{eq:constant}.

We next discuss how the appearance of the extra constants $C^{\mathcal{B}}_{ab}$ and $C_{bac}$ defined in \eqref{eq:constCb} and ~\eqref{Cbac} affect the boundary crossing relation. For this purpose, we introduce auxiliary boundary S-matrices $\widetilde{R}^{a}_{\, bc}$ and $\widetilde{\mathsf{R}}^{a}_{\, bc}$, which are obtained by incorrectly amputating two-point functions \eqref{eq:two_pt} and \eqref{eq:two_pt_rotated} by correlators \eqref{eq:wrong_amputate} and \eqref{wrongamputatefunction_bdy}, respectively. As discussed in \cite{Ghoshal:1993tm}, these two boundary S-matrices are subject to the following equation:
\begin{align}\label{eq:cross_bdry}
    \widetilde{R}^{a}_{\, bc}(\theta)=\widetilde{\mathsf{R}}^{a}_{\, bc}(\theta)\ . 
\end{align}
However, as in the bulk case, the correctly normalized boundary S-matrices $R^{a}_{\, bc}$ and $\mathsf{R}^{a}_{\, bc}$ should be amputated by the correlation functions \eqref{eq:correct_amputate} and \eqref{correctamputatefunction_bdy}. By using the relations \eqref{eq:constCb} and \eqref{Cbac}, it turns out that the correctly normalized boundary S-matrices are related to the incorrectly normalized ones as
\begin{align}
    R^{a}_{\, bc}(\theta)=\frac{1}{\sqrt{C^{\mathcal{B}}_{ab}\cdot C^{\mathcal{B}}_{ac}}}\, \widetilde{R}^{a}_{\, bc}(\theta)\, , \quad \mathsf{R}^{a}_{\, bc}(\theta)=\frac{1}{\sqrt{C_{bac}}}\, \widetilde{\mathsf{R}}^{a}_{\, bc}(\theta)\ . 
\end{align}
By plugging these expressions into \eqref{eq:cross_bdry}, one can obtain the precise relation between boundary S-matrices in different pictures
\begin{align}\label{eq:precise_crossing_relation}
    \mathsf{R}^{a}_{\, bc}(\theta)=\sqrt{\frac{C^{\mathcal{B}}_{ab}\cdot C^{\mathcal{B}}_{ac}}{C_{bac}}}\, R^{a}_{\, bc}(\theta)\ . 
\end{align}
As mentioned in Sec.~\ref{subsec:boundary_S_matrix}, in the rotated picture, the kink state belongs to the bulk Hilbert space, hence in and out states are related via the unitary transformation implemented by the correctly normalized bulk S-matrix $S^{ab}_{dc}$:
\begin{align}
    {\mathsf{R}}^{b}_{\, ca}\left(\frac{\i\pi}{2}-\theta\right)&=\sum_{d}{S}^{ba}_{cd}(2\theta)\, {\mathsf{R}}^{d}_{\, ca}\left(\frac{\i\pi}{2}+\theta\right)\ .
\end{align}
By substituting \eqref{eq:precise_crossing_relation} into this, one can arrive at the boundary crossing relation:
\begin{align}\label{eq:boundary_crossing_int}
    R^{b}_{\, ca}\left(\frac{\i\pi}{2}-\theta\right)&=\sum_{d}\sqrt{\frac{\d_{d}}{\d_{b}}\cdot \frac{C^{\mathcal{B}}_{dc}\cdot C^{\mathcal{B}}_{da}}{C^{\mathcal{B}}_{bc}\cdot C^{\mathcal{B}}_{ba}}}\, {S}^{ba}_{cd}(2\theta)\, R^{d}_{\, ca}\left(\frac{\i\pi}{2}+\theta\right)\ , 
\end{align}
where we used the expression \eqref{eq:constant}.

All that remains is to discuss the constant $C^{\mathcal{B}}_{ab}$. 
The fusion property \eqref{eq:fusionproperty} allows us to evaluate the left diagram in \eqref{eq:constCb} as follows:
\begin{align}
        \begin{tikzpicture}
        \pgfmathsetmacro{\radius}{1.5/sqrt(2)}
        \begin{scope}[scale=0.8, shift={(-2cm, 0)}]
        \node at ($(3.7-8-0.2,1+0.2)$) {$b$}; 
        \node at ($(4-8-0.3,0) + (180:1)$) {$a$}; 
        \node at ($(3.7-8-0.2,-1-0.2)$) {$b$};
        \fill[black!20!white] ($(4-8, 1.914)$)--($(4.5-8, 1.914)$)--($(4.5-8, -1.914)$)--($(4-8, -1.914)$);
        \draw[thick, blue!80] ($(-4, 1.5)$) arc (90:270: 1.5cm);
        \draw[thick, black!20!green] ($(-4,0)+(225:1.5)$) arc (-90:90: 0.5cm and \radius cm);
                 \foreach \angle in {135, 225} {
             \fill[black] ($(-4,0)+(\angle:1.5)$) circle(2pt);
        \draw[purple, ultra thick] (-4, 1.914)--(-4, -1.914);
         }
        \end{scope}
        \node[right] at (-4,0) {$=$}; 
        \node[right] at (-3,0) {{$\sum\limits_{d\in b \otimes b}\, \frac{\d_{d}}{\d_{b}}$}};
        \begin{scope}[scale=0.8, shift={(-8.8cm, 0)}]
        \draw[thick, black!20!green] (8.5,-1) -- (8.5, 0);
        \draw[thick, black!20!green] (8.5, 0) -- (8.5, 1);
        \draw [blue!80, thick] (8.5,0) circle (1);
        \draw [blue!80, thick] (9.5,0)--(10.5, 0);
        \fill[blue!80] (9.5, 0) circle (2pt);
        \fill[blue!80] (10.5, 0) circle (2pt);
        \node[below] at (10, 0) {$d$};
        \fill[black] (8.5, 1) circle (2pt);
        \fill[black] (8.5, -1) circle (2pt);
        \node at ($(8,0)$) {$a$}; 
        \node at ($(9,0.5)$) {$b$};
        \node at ($(9,-0.5)$) {$b$};
        \node at ($(11,0.5)$) {$b$};
        \node at ($(11,-0.5)$) {$b$};
        \draw [blue!80, thick] (11.5,1) arc(90:270:1cm);
        \fill[black!20!white] ($(11.5, 1.914)$)--($(12, 1.914)$)--($(12, -1.914)$)--($(11.5, -1.914)$);
        \draw[purple, ultra thick] (11.5, 1.914)--(11.5, -1.914);
        \end{scope}
        \node[right] at (3.5,0) {$=$};
        \begin{scope}[scale=0.8, shift={(-0.8cm, 0)}]
        \node[right] at (6.2, 0) {{$\frac{1}{\d_{b}}$}};
        \draw[thick, black!20!green] (8.5,-1) -- (8.5, 0);
        \draw[thick, black!20!green] (8.5, 0) -- (8.5, 1);
        \draw [blue!80, thick] (8.5,0) circle (1);
        \fill[black] (8.5, 1) circle (2pt);
        \fill[black] (8.5, -1) circle (2pt);
        \node at ($(8,0)$) {$a$}; 
        \node at ($(9,0)$) {$b$};
        \node at ($(11,0)$) {$b$};
        \draw [blue!80, thick] (11.5,1) arc(90:270:1cm);
        \fill[black!20!white] ($(11.5, 1.914)$)--($(12, 1.914)$)--($(12, -1.914)$)--($(11.5, -1.914)$);
        \draw[purple, ultra thick] (11.5, 1.914)--(11.5, -1.914);
        \end{scope}
        \end{tikzpicture}
\end{align}
In the final equality, we used the fact that the two kinks are sufficiently far from the boundary, and the only identity line contributes effectively. (See also footnote \ref{footnote}.) Moreover, we define
\begin{align}
    \begin{tikzpicture}[scale=0.8, baseline=0cm]
        \draw [blue!80, thick] (11.5,1) arc(90:270:1cm);
        \fill[black!20!white] ($(11.5, 1.914)$)--($(12, 1.914)$)--($(12, -1.914)$)--($(11.5, -1.914)$);
        \draw[purple, ultra thick] (11.5, 1.914)--(11.5, -1.914);
        \node at ($(11,0)$) {$b$};
        \node at (9, 0) {$\frac{1}{F^{\mathcal{B}}_{b}}\ \ = $};
    \end{tikzpicture}
\end{align}
where $F^{\mathcal{B}}_{b}$ is some constant which depends on details of the topological line $\mathcal{L}_{b}$ and the boundary $\mathcal{B}$. By putting the above discussions together, it turns out that the constant ${C}^{\mathcal{B}}_{ab}$ can be written as
\begin{align}\label{eq:constB}
    {C}^{\mathcal{B}}_{ab}=\frac{1}{\d_{b}\cdot F^{\mathcal{B}}_{b}}\ .
\end{align}
The crucial point here is that the constant ${C}^{\mathcal{B}}_{ab}$ does {\it not} depend on the topological line $\mathcal{L}_{a}$, which leads to
\begin{align}
    \frac{C^{\mathcal{B}}_{dc}\cdot C^{\mathcal{B}}_{da}}{C^{\mathcal{B}}_{bc}\cdot C^{\mathcal{B}}_{ba}}=1\ . 
\end{align}
By plugging this result into \eqref{eq:boundary_crossing_int}, we can obtain the precise boundary crossing relation: 
\begin{align}
    {R}^{b}_{\, ca}\left(\frac{\i\pi}{2}-\theta\right)&=\sum_{d}\sqrt{\frac{\d_d}{\d_b}}\, {S}^{ba}_{cd}(2\theta)\, {R}^{d}_{\, ca}\left(\frac{\i\pi}{2}+\theta\right)\ .
\end{align}
This is the main result of this paper. We note that this boundary crossing relation differs from the conventional form derived by Ghoshal and Zamolodchikov due to the presence of the TQFT factor $\sqrt{\d_{d}/\d_{b}}$.

For future reference, we close this section by summarizing the consistency conditions to which the boundary S-matrix ${R}^{a}_{\,bc}$ must be subject:
\begin{itemize}
    \item Boundary unitarity condition:
        \begin{align}\label{eq:bdy_unitary_new}
            \sum_{d}{R}^{b}_{\, cd}(\theta){R}^{b}_{\, da}(-\theta)=\delta_{ac}\ , 
        \end{align}
    \item Boundary crossing relation:
        \begin{align}\label{eq:bdy_crossing_relation_new}
    {R}^{b}_{\, ca}\left(\frac{\i\pi}{2}-\theta\right)&=\sum_{d}\sqrt{\frac{\d_d}{\d_b}}\, {S}^{ba}_{cd}(2\theta)\, {R}^{d}_{\, ca}\left(\frac{\i\pi}{2}+\theta\right)\ ,
\end{align}
    \item Boundary Yang-Baxter equation:
    \begin{align}\label{eq:bdy_YB_new}
    \begin{aligned}
        &\sum_{f, g}{S}^{g a}_{c b}(\theta_1 -\theta_2){R}^{g}_{\,af}(\theta_1 ){S}^{d f}_{c g}(\theta_1 +\theta_2){R}^{d}_{\,fe}(\theta_2 ) \\
        &\qquad \qquad =\sum_{f, g} {R}^{b}_{\, ag}(\theta_2 ) {S}^{f g}_{c b}(\theta_1 +\theta_2 ){R}^{f}_{\, ge}(\theta_1 ) {S}^{de}_{c f}(\theta_1 -\theta_2 )\ . 
    \end{aligned}
\end{align}
\end{itemize}

\section{Boundary Scattering in Gapped Theories from Minimal Models}\label{sec:boundary_scattering_example}
In the previous section, we saw that the correctly normalized boundary S-matrix taken into account the TQFT dynamics brings about the modification of the boundary crossing relation. In this section, we demonstrate how this modified boundary crossing relation can be used for deriving the exact boundary S-matrix with a non-invertible symmetric boundary in conjunction with boundary unitarity condition and Yang-Baxter equation. 
To be concrete, we particularly focus on the boundary scattering where the bulk theory is given by the gapped theory obtained by the relevant deformation of the minimal model $M_{p}$ ($p\geq 4$) discussed in Sec.~\ref{subsec:IR_minimal}. One can generally put the boundary S-matrix into the following form:
\begin{align}\label{eq:general_boundary_S_matrix}
    {R}^{a}_{\, bc}(\theta)=\delta_{b\not=c}\, X^{a}_{\, bc}(\theta)+\delta_{bc}\left(\delta_{b, a+1}\, U^{a}_{\, bb}(\theta)+\delta_{b, a-1}\, U^{a}_{\, bb}(\theta)\right) \ . 
\end{align}
While the first part $X^{a}_{\, bc}$ describes the {\it non-diagonal} scattering where an initial kink is scattered to a different kink, the second part $U^{a}_{\, bb}$ represents the {\it diagonal} scattering where initial and final kinks are identical. 
In what follows, we determine these scattering amplitudes under the assumption that the boundary preserves the categorical symmetry $\mathcal{A}_{p}$ along with the bulk RG flow\footnote{For instance, if we choose the UV boundary state as the orbit of the categorical symmetry $\mathcal{A}_{p}$, i.e., 
\begin{align*}
    \ket{\mathcal{B}}=\sum_{c}\ket{\mathcal{L}_c }\ ,
\end{align*}
where $\ket{\mathcal{L}_c }$ is the Cardy state associated to the topological line defect $\mathcal{L}_{c}$, the boundary is trivially weakly-symmetric under the fusion category $\mathcal{A}_{p}$.
}.
\subsection{Ward-Takahashi Identities for Non-invertible Symmetries}
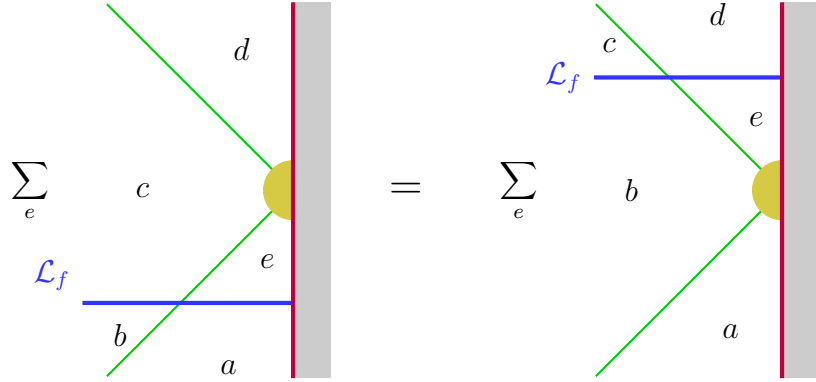
\begin{figure}
    \centering
            \begin{tikzpicture}
        \fill[black!20!white] (4,2.5)--(4,-2.5)--(4.5, -2.5)--(4.5, 2.5)--cycle;
        \fill[black!20!white] (10.5,2.5)--(10.5,-2.5)--(11, -2.5)--(11, 2.5)--cycle;
        \node  at (0.5,0) {$\displaystyle \sum_{e}$}; 
        \foreach \angle in {135, 225} {
            \draw[thick, black!20!green] (4,0) -- ++(\angle:3.5);
        }
        \draw[blue!80, ultra thick] (1.2,-1.5) node[above left]{$\CL_{f}$} --(4,-1.5);
        \fill[black!20!yellow] (4,0.4) arc[start angle=90, end angle=270, radius=0.4cm] -- cycle;
        \draw[purple, ultra thick] (4,2.5)--(4,-2.5);
        \node at ($(4,0) + (110:2)$) {$d$}; 
        \node at ($(4,0) + (180:2)$) {$c$}; 
        \node at ($(4,0) + (250:1)$) {$e$};
        \node at ($(4,0) + (250:2.5)$) {$a$};
        \node at ($(4,0) + (220:3)$) {$b$};
        \node at (5.5,0) {{\Large $=$}}; 
        \node at (7,0) {$\displaystyle \sum_{e}$}; 
        \foreach \angle in {135, 225} {
            \draw[thick, black!20!green] (10.5,0) -- ++(\angle:3.5);
        }
        \draw[blue!80, ultra thick] (8,1.5) node[left]{$\CL_{f}$} --(10.5,1.5);
        \fill[black!20!yellow] (10.5,0.4) arc[start angle=90, end angle=270, radius=0.4cm] -- cycle;
        \draw[purple, ultra thick] (10.5,2.5)--(10.5,-2.5);
        \node at ($(10.5,0) + (110:2.5)$) {$d$}; 
        \node at ($(10.5,0) + (180:2)$) {$b$}; 
        \node at ($(10.5,0) + (110:1)$) {$e$};
        \node at ($(10.5,0) + (250:2)$) {$a$};
        \node at ($(10.5,0) + (140:3)$) {$c$};
\end{tikzpicture}
    \caption{Ward-Takahashi identities associated with the categorical symmetry $\mathcal{A}_{p}$.}
    \label{fig:WT_id_boundary}
\end{figure}

We begin by analyzing the constraints on the boundary S-matrices $X^{a}_{\, bc}$ and $U^{a}_{\, bb}$ from the categorical symmetry $\mathcal{A}_{p}$. The Ward-Takahashi identities associated to  $\mathcal{A}_p$ can be written as (see also Fig.\,\ref{fig:WT_id_boundary}.)
\begin{align}\label{eq:Ward_Takahashi_bdry_generalp}
    \begin{aligned}
        \mathsf{WT}(\mathcal{L}_{f}):=\sum_{e\in f\otimes a} \delta_{e, c\pm1}{R}^{c}_{\, ed}(\theta)
        -\sum_{e\in f\otimes d} \delta_{e, b\pm1}{R}^{b}_{\, ae}(\theta)=0\ , 
    \end{aligned}
\end{align}
where $\delta_{b, a\pm1}:=\delta_{b, a+1}+\delta_{b, a-1}$.
At first glance, these constraints seem to be difficult to solve, it is sufficient to consider the Ward-Takahashi identity only for $f=2$. To see this, we show the following statement:
\begin{align}
    \mathsf{WT}(\mathcal{L}_{2})=0\qquad \Longrightarrow \qquad \mathsf{WT}(\mathcal{L}_{f})=0 \quad (\forall f=1,2,\cdots p-1)\ .  
\end{align}
\paragraph{Proof.} We prove this by mathematical induction. For $f = 1$ and $f = 2$, the Ward–Takahashi identities are trivially satisfied. Now, assume that for $g = 3, \dots, p-1$, the following Ward–Takahashi identities hold:
\begin{align}\label{eq:assumption}
    \mathsf{WT}(\mathcal{L}_{h})=0\ , \quad h<g\ .
\end{align}
We will show that $\mathsf{WT}(\mathcal{L}_{g}) = 0$.
To this end, consider $\mathsf{WT}(\mathcal{L}_2 \otimes \mathcal{L}_{g-1})$.
By successively moving the topological lines $\mathcal{L}_2$ and $\mathcal{L}_{g-1}$ across the boundary S-matrix, and using the inductive assumption \eqref{eq:assumption}, we find:
\begin{align}\label{eq:wt2g2}
    \mathsf{WT}(\mathcal{L}_{2}\otimes\mathcal{L}_{g-1})=0\ , 
\end{align}
Furthermore, the fusion property \eqref{eq:fusionproperty} implies that for $g \geq 3$, the following fusion rule holds:
\begin{align}
    \mathcal{L}_{2}\otimes\mathcal{L}_{g-1}=\mathcal{L}_{g-2}\oplus \mathcal{L}_{g}\ . 
\end{align}
By definition, the function $\mathsf{WT}$ is additive under direct sums of topological lines, i.e., $\mathsf{WT}(\mathcal{L}_{i}\oplus \mathcal{L}_{j})=\mathsf{WT}(\mathcal{L}_{i})+\mathsf{WT}(\mathcal{L}_{j})$. These two properties allow us to rewrite \eqref{eq:wt2g2} as
\begin{align}
    0\overset{\eqref{eq:wt2g2}}{=}\mathsf{WT}(\mathcal{L}_{2}\otimes\mathcal{L}_{g-1})=\mathsf{WT}(\mathcal{L}_{g-2})+\mathsf{WT}(\mathcal{L}_{g})=\mathsf{WT}(\mathcal{L}_{g})\ ,
\end{align}
where in the last equality, we again used \eqref{eq:assumption}. This completes the proof.

Therefore, it suffices to consider only $f=2$ in \eqref{eq:Ward_Takahashi_bdry_generalp}, and the boundary S-matrix is constrained as follows depending on the value of $p$:
\begin{itemize}
    \item $p=4$ 
        \begin{align}
            R(\theta)&:=U^{1}_{22}(\theta)=U^{3}_{22}(\theta)\ , \\
             P(\theta)&:=U^{2}_{33}(\theta)=U^{2}_{11}(\theta)\ , \\
            V(\theta)&:=X^{2}_{31}(\theta)=X^{2}_{13}(\theta)\ , \\
            R(\theta)&=P(\theta)+V(\theta)\ . \label{eq:non-inv_const;p=4} 
        \end{align}
    \item $p\geq 5$
        \begin{align}\label{eq:non_inv_analysis:p>4}
            X^{a}_{bc}(\theta)=0\ , \quad U(\theta):=U^{a}_{bc}(\theta)\ , \quad \forall a, b, c\ . 
        \end{align}
\end{itemize}
Interestingly, we find that the non-diagonal scatterings are prohibited and the all diagonal scatterings become to the identical one for $p\geq 5$ without relying on the bulk and boundary integrability. 
We next derive the exact boundary S-matrix for $p=4$ and $p\geq 5$ in order.
\subsection{$p=4$}
Here, we consider the boundary scattering for $p=4$, where the corresponding UV minimal model is the so-called tri-critical Ising model $M_4$. In this section, we assume the boundary integrability, which allows us to use the boundary Yang-Baxter equation \eqref{eq:bdy_YB_new}, and examine its implications for boundary scattering.
\subsubsection*{Constraints from boundary Yang-Baxter equation}
The boundary Yang-Baxter equation \eqref{eq:bdy_YB_new} gives the following constraint:
\begin{align}\label{eq:bdy_YB_tri_2_new}
            \begin{aligned}
                &(1+\sqrt{2}f_{+})V(\theta_1 )P(\theta_2 )+P(\theta_1 )V(\theta_2 )\\
                &\qquad =(1+\sqrt{2}f_{+})(1+\sqrt{2}f_{-})P(\theta_1 )V(\theta_2 )+(1+\sqrt{2}f_{-})V(\theta_1 )P(\theta_2 )\ ,
            \end{aligned}
        \end{align}
where $f_{\pm}$ is defined by
        \begin{align}
            f_{\pm}:=\frac{\sinh\left(\frac{\theta_{1}\pm\theta_{2}}{4}\right)}{\sinh\left(\frac{\i\pi-\theta_{1}\mp\theta_{2}}{4}\right)}\ . 
        \end{align}
Suppose that the diagonal scattering amplitude does not vanish, namely $P(\theta)\not= 0$, we introduce 
\begin{align}
    Q(\theta):=\frac{V(\theta)}{P(\theta)}\ .
\end{align}
This simplifies \eqref{eq:bdy_YB_tri_2_new} to
\begin{align}
     \begin{aligned}
                &(1+\sqrt{2}f_{+})Q(\theta_1 )+Q(\theta_2 ) =(1+\sqrt{2}f_{+})(1+\sqrt{2}f_{-})Q(\theta_2 )+(1+\sqrt{2}f_{-})Q(\theta_1 )\ .
            \end{aligned}
\end{align}
By taking the limit $\theta_{2}\to \theta_1$, this equation turns into the ordinary differential equation:
\begin{align}
    \frac{\text{d} Q(\theta)}{\text{d}\theta}-\frac{1}{2\tanh(\theta/2)}Q(\theta)=0\ , 
\end{align}
We can readily solve this differential equation, and relate the boundary S-matrices $P(\theta)$ and $V(\theta)$ as follows
\begin{align}\label{eq:eq1}
    \frac{V(\theta)}{P(\theta)}=A \sinh\left(\frac{\theta}{2}\right)\ . 
\end{align}
where $A$ is a free parameter. Moreover, by plugging this into the constraint from the non-invertible symmetry \eqref{eq:non-inv_const;p=4}, we can also relate $P(\theta)$ and $R(\theta)$ as
\begin{align}\label{eq:R}
    \frac{R(\theta)}{P(\theta)}=1+A\sinh\left(\frac{\theta}{2}\right)\ .
\end{align}
\subsubsection*{Constraints from the boundary crossing relation}
We next move on to the constraints coming from the boundary crossing relation \eqref{eq:bdy_crossing_relation_new}. We should notice that there are no any modifications in the boundary crossing relation when $p=4$. (As we will see in the next section, non-trivial modifications start from $p=5$.) By setting $(a,b,c)=(3, 2, 3)$ in the boundary crossing relation \eqref{eq:bdy_crossing_relation_new}, we obtain the following constraint:
\begin{align}
    P\left(\frac{\i\pi}{2}-\theta\right)=Z_{4}(2\theta)\,\sinh\left(\frac{\text{i}\pi+2\theta}{4}\right)P\left(\frac{\i\pi}{2}+\theta\right)\ , \label{eq:bdry_crosing_+0+}
\end{align}
where $Z_{4}(\theta)$ is given by \eqref{eq:explicit_form_Z}.
Also, the boundary crossing relation for $(a, b, c)=(2, 3, 2)$ gives 
\begin{align}\label{eq:crossing_rel}
     R\left(\frac{\i\pi}{2}-\theta\right)=Z_{4}(2\theta)\,\sinh\left(\frac{\text{i}\pi+2\theta}{4}\right)R\left(\frac{\i\pi}{2}+\theta\right)\ .
\end{align}
Keeping in mind that we now assume $P(\theta)\not =0$, these two boundary crossing relations give rise to the following constraint:
\begin{align}
    \frac{R\left(\frac{\i\pi}{2}-\theta\right)}{P\left(\frac{\i\pi}{2}-\theta\right)}=\frac{R\left(\frac{\i\pi}{2}+\theta\right)}{P\left(\frac{\i\pi}{2}+\theta\right)}\ .
\end{align}
By plugging the expression \eqref{eq:R} into this, one can arrive at
\begin{align}
    A\sinh\left(\frac{\i \pi-2\theta}{4}\right)=A\sinh\left(\frac{\i \pi+2\theta}{4}\right)\ . 
\end{align}
This is the identity with respect to the rapidity $\theta$, hence the free parameter $A$ is forced to be zero. Thereby, recalling the relation \eqref{eq:eq1}, we can conclude that if we assume that the diagonal scattering $P(\theta)$ is non-trivial, the non-diagonal scattering must be trivial:
\begin{align}
    P(\theta)\not=0 \quad \Longrightarrow \quad V(\theta)=0\ . 
\end{align}
Similarly, one can also show that if we assume that the non-diagonal scattering is non-trivial, the diagonal scattering $P(\theta)$ must be trivial:
\begin{align}\label{eq:converse}
    V(\theta)\not=0 \quad \Longrightarrow \quad P(\theta)=0\ . 
\end{align}
By combining these results with \eqref{eq:R} and its counterpart for the second case \eqref{eq:converse}, the discussions above can be summarized as follows. When the boundary is weakly-symmetric under $\mathcal{A}_4$, and the bulk and boundary integrability hold, one can consider the following two scenarios of the boundary scattering:
\begin{align}
    \text{Scenario 1 : } R(\theta)=P(\theta)\not =0\ , \quad V(\theta)=0\ , \\
    \text{Scenario 2 : } R(\theta)=V(\theta)\not =0\ , \quad P(\theta)=0\ .
\end{align}
Finally, by combining the boundary crossing relation \eqref{eq:crossing_rel} with the following boundary unitarity condition:
\begin{align}\label{eq:uni_p=4}
    R(\theta)R(-\theta)=1\ ,
\end{align}
one can obtain the minimal solution:\footnote{We should mention that the same equations as \eqref{eq:crossing_rel} and \eqref{eq:uni_p=4} can also be seen in \cite{Ghoshal:1993tm} where the boundary scattering in the sine-Gordon model is discussed. We adopt their solution in our discussion. This remark is also valid for the $p\geq 5$ case. (See the discussions around \eqref{eq:boundary_S_matrix_p>4}.)}
\begin{align}\label{eq:boundary_S_matrix_p=4}
    R(\theta)= \frac{\widetilde{F}_{4}(\theta)}{\widetilde{F}_{4}(-\theta)}\ , 
\end{align}
where $\widetilde{F}_{p}(\theta)$ is given by
\begin{align}\label{eq:F}
    \widetilde{F}_{p}(\theta)=\frac{\Gamma\left(1+\frac{2\i\theta}{p\pi}\right)}{\Gamma\left(\frac{1}{p}+\frac{2\i\theta}{p\pi}\right)}\prod_{k=1}^{\infty}\frac{\Gamma\left(\frac{4k}{p}+\frac{2\i\theta}{p\pi}\right)\Gamma\left(1+\frac{4k}{p}+\frac{2\i\theta}{p\pi}\right)\Gamma\left(\frac{4k+1}{p}\right)\Gamma\left(1+\frac{4k-1}{p}\right)}{\Gamma\left(\frac{4k+1}{p}+\frac{2\i\theta}{p\pi}\right)\Gamma\left(1+\frac{4k-1}{p}+\frac{2\i\theta}{p\pi}\right)\Gamma\left(1+\frac{4k}{p}\right)\Gamma\left(\frac{4k}{p}\right)}\ . 
\end{align}
\subsection{$p\geq 5$}
Unlike the $p=4$ case, the non-diagonal scatterings are automatically prohibited from the Ward-Takahashi identities associated to the categorical symmetry $\mathcal{A}_{p}$. We can thus focus on the diagonal scatterings from the beginning. Notably, the non-trivial modifications due to the TQFT dynamics to the boundary crossing relations start from $p=5$. This time, the modified boundary crossing relation \eqref{eq:bdy_crossing_relation_new} reads 
\begin{align}
    U\left(\frac{\i\pi}{2}-\theta\right)&=\sum_{d=a\pm 1}\sqrt{\frac{\d_d}{\d_b}}\, {S}^{ba}_{ad}(2\theta)\, U\left(\frac{\i\pi}{2}+\theta\right)\ .
\end{align}
Surprisingly, it turns out that the summation does not rely on the vacuum indices $a$ and $b$:
\begin{align}
    \sum_{d=a\pm 1}\sqrt{\frac{\d_d}{\d_b}}\, {S}^{ba}_{ad}(2\theta)=Z_{p}(2\theta)\sinh\left(\frac{\i\pi+2\theta}{p}\right)\ . 
\end{align}
This can be checked by using the explicit form of the quantum dimension \eqref{eq:quantum_dimension} and the bulk S-matrix \eqref{eq:S_matrix_RSOS_new}. Then, the modified boundary crossing relation can be simply put as
\begin{align}
    U\left(\frac{\i\pi}{2}-\theta\right)&=Z_{p}(2\theta)\sinh\left(\frac{\i\pi+2\theta}{p}\right)\, U\left(\frac{\i\pi}{2}+\theta\right)\ .
\end{align}
We should emphasize that we have not used the boundary integrability up to here.

In a similar way to the $p=4$ case, by combining this with the boundary unitarity condition
\begin{align}
        U(\theta) U(-\theta)=1\ ,
\end{align}
one can  derive the boundary S-matrix which is consistent with the Ward-Takahashi identities for non-invertible symmetries:
\begin{align}\label{eq:boundary_S_matrix_p>4}
    U(\theta)=\frac{\widetilde{F}_{p}(\theta)}{\widetilde{F}_{p}(-\theta)}\ , 
\end{align}
where $\widetilde{F}_{p}(\theta)$ is given in \eqref{eq:F}.

\section{Conclusion and Future Prospects}\label{sec:conclusion}
In this paper, we revisited boundary scattering in $1+1$ dimensions with weakly-symmetric boundaries under the non-invertible symmetries. Our analysis was based on the assumption that a categorical symmetry of the UV theory remains unbroken along the bulk RG flow, and is spontaneously broken in the infrared regime. Within this framework, we investigated the scattering of kinks, which are characterized by the categorical symmetry.

One of our key findings is that the boundary crossing relation proposed by Ghoshal and Zamolodchikov undergoes a nontrivial modification due to the effects of the TQFT dynamics associated with the spontaneous breaking of the non-invertible symmetry. By paying close attention to the normalization of the boundary S-matrix, we derived the precise boundary crossing relation \eqref{eq:bdy_crossing_relation_new}. 

As a concrete example, we considered a setup where the UV theory is the minimal model $M_{p}$ $(p\geq 4)$, deformed by the $\Phi_{(1,3)}$-deformation, which induces a bulk RG flow to the gapped theory. Throughout the flow, the categorical symmetry $\mathcal{A}_p$ whose fusion rule is given by \eqref{fusion_rule} is preserved, allowing us to describe the theory in terms of $(p-2)$-kinks labeled by this global symmetry. We analyzed the boundary scattering amplitudes for these kinks based on the modified boundary crossing relation we derived. 
First, we investigated the constraints on boundary scatterings imposed by the Ward-Takahashi identities associated to the categorical symmetry. In particular, we found that for $p\geq5$, the symmetry considerations alone are sufficient to prohibit non-diagonal scattering and to enforce identical diagonal scattering for all kinks. 
Finally, combining these symmetry constraints with the three consistency conditions, we obtained the minimal solution of the boundary S-matrix. 

To conclude this paper, we highlight several intriguing directions for future research.
\begin{itemize}
\item First, in our analysis, we primarily focused on the bulk $\Phi_{(1,3)}$-deformation as a concrete example. It would be interesting to explore the boundary scattering in gapped theories obtained from other relevant deformations. For instance, in the case of the tri-critical Ising model deformed by $\Phi_{(2,1)}$, it is known that bulk integrability and the Fibonacci categorical symmetry are preserved \cite{Zamolodchikov:1989hfa, Zamolodchikov:1990xc, Chang:2018iay}, and that the spectrum consists not only of kinks but also of the breather. A natural extension of our work would be to apply our modified boundary crossing relation to these excitations and investigate the boundary S-matrices among them. Additionally, beyond bulk deformations, it would be valuable to study boundary RG flows and how they affect the boundary scattering.

\item Another potential direction is the study of monopole scattering. Consider a four-dimensional scattering where a fermion is scattered off a very massive monopole. It has been argued that, when restricted to the s-wave sector, this problem effectively reduces to a two-dimensional boundary scattering problem \cite{Polchinski:1984uw, Maldacena:1995pq, Brennan:2021ewu, Hamada:2022eiv, vanBeest:2023dbu, vanBeest:2023mbs, Loladze:2024ayk}. Investigating how the global symmetry influences the boundary scattering in this context could provide new insights into the interplay between symmetry and monopole dynamics.

\item Furthermore, in this paper, we assumed that vacuum states are  objects in the regular module category associated to the categorical symmetry. A natural question is how our boundary crossing relation would be modified if the vacuum instead belongs to a non-regular module category. In this regard, the framework developed in \cite{Copetti:2024dcz}, where the symmetry TFT was used to study modifications of the bulk S-matrix crossing relation, may offer useful insights for addressing this problem.

\item Finally, it is much intriguing to apply our result to the boundary bootstrap. Recently, the S-matrix bootstrap is extended to include the boundary QFTs in $1+1$ dimensions~\cite{Kruczenski:2020ujw}. In particular, the authors in \cite{Kruczenski:2020ujw} clarify that the constraints coming form the analyticity, boundary crossing relation and unitarity can be used to bound the possible boundary S-matrix by using the bosonic $\text{O}(N)$ vector model. It is worth investigating how the modified boundary crossing relation derived in this paper influences the bootstrap analysis of the boundary S-matrix.
\end{itemize}

\section*{Acknowledgement}
We thank Yuta Hamada, Shota Komatsu and Yifan Wang for helpful discussions. The authors thank Kyushu University Institute for Advanced Study and RIKEN Interdisciplinary Theoretical and Mathematical Sciences Program. Discussions during the ``Kyushu IAS-iTHEMS workshop: Non-perturbative methods in QFT'' were useful in completing this work. S.\,S. is supported by Grant-in-Aid for JSPS Fellows No.~23KJ1533. S.\,Y. is supported by JSPS KAKENHI No.~21K03574.

\bibliographystyle{utphys}
\bibliography{bdry_scattering}

\end{document}